\documentclass[usenatbib]{mn2e}
\usepackage[utf8]{inputenc}
\usepackage{natbib,epsfig,amsmath,cite,amssymb,graphicx,mystyle}
\usepackage{mathtools}
\usepackage{array, xcolor, caption, subcaption, multirow, multicol, ulem}
\usepackage[colorlinks=true,linkcolor=blue,citecolor=blue]{hyperref}

\title[R-process enrichment by binary neutron star mergers]{
The impact of turbulent mixing on the galactic r-process enrichment by binary neutron star mergers
}

\author[Dvorkin et al.]{Irina Dvorkin$^{1}$\thanks{E-mail: dvorkin@iap.fr}, Fr\'{e}d\'{e}ric Daigne$^{1}$, Stephane Goriely$^2$, Elisabeth Vangioni$^{1}$, \newauthor Joseph Silk$^{1,3,4}$\\
$^{1}$ Institut d'Astrophysique de Paris, Sorbonne Universit\'{e} \& CNRS, UMR 7095, 98 bis bd Arago,\\ F-75014 Paris, France\\
$^{2}$ Institut d'Astronomie et d'Astrophysique, Universit\'{e} Libre de Bruxelles, Campus de la Plaine, CP-226, 1050 Brussels, Belgium \\
$^{3}$ Department of Physics and Astronomy, The Johns Hopkins University,
Baltimore, MD 21218, USA \\
$^{4}$ BIPAC, University of Oxford, 1 Keble Road, Oxford OX1 3RH, UK \\
}

\begin{document}
\newcommand{\dd}{{\rm d}}

\pagerange{\pageref{firstpage}--\pageref{lastpage}} \pubyear{2020}
\maketitle
\label{firstpage}

\begin{abstract}
We study the enrichment of the interstellar medium with rapid neutron capture (r-process) elements produced in binary neutron star (BNS) mergers. We use a semi-analytic model to describe  galactic evolution, with merger rates and time delay distributions of BNS mergers consistent with the latest population synthesis models. In order to study the dispersion of the relative abundances of r-process elements and iron, we applied a turbulent mixing scheme, where the freshly synthesized elements are gradually dispersed in the interstellar medium. We show that within our model the abundances observed in Milky-Way stars, in particular the scatter at low metallicities, can be entirely explained by BNS mergers. Our results  suggest that binary neutron star mergers could be the dominant source of r-process elements in the Galaxy.
\end{abstract}


\section{Introduction}

The origin of rapid neutron-capture process (r-process) elements is a long-standing question, on the crossroads between nuclear astrophysics, stellar evolution and galaxy formation \citep{Arnould07,2019arXiv190101410C,Arnould20}. R-process nucleosynthesis, through which about half of all elements above the iron group are produced, requires extreme neutron densities, which can only exist in cataclysmic astrophysical environments. The two astrophysical events considered as the main potential r-process sites are (i) binary neutron star (BNS) or neutron star-black hole mergers \citep[e.g.][]{1974ApJ...192L.145L,1999A&A...341..499R,Goriely11b,Just15} and (ii) rare core-collapse supernovae (SNe), such as collapsars or forming magnetars \citep[e.g.][]{1985ApJ...291L..11S,2018MNRAS.477.2366H,2019Natur.569..241S}. Both these classes of events can, in principle, produce enough r-process elements to explain the abundances observed in the atmospheres of Milky-Way (MW) stars, and occur at rates that are at most $\sim 1\%$ of the regular core-collapse SN rate, which may be the origin of the observed scatter (up to two orders of magnitude) of r-process relative to iron-group element abundances.

The detection of gravitational waves (GW) from the GW170817 BNS merger and the subsequent identification of its electromagnetic counterparts \citep{2017PhRvL.119p1101A,2017ApJ...848L..12A,2017Sci...358.1556C} have provided strong support to the first hypothesis.  The optical/infrared emission is identified as a kilonova, powered by the  decay heat from r-process nuclei produced in the merger ejecta \citep{2017ApJ...848L..12A,2017ApJ...848L..19C,2017Sci...358.1570D,2017Sci...358.1583K,2017Natur.551...67P,2017Sci...358.1574S}. The high opacities and the total ejected mass of $\sim 0.05M_{\odot}$ inferred from the observed spectral evolution of this event are consistent with model predictions \citep{1998ApJ...507L..59L,2010MNRAS.406.2650M,2017Natur.551...80K}. Moreover, r-process nucleosynthesis in BNS mergers was shown to produce a solar abundance pattern \citep{Goriely11b,2013ApJ...773...78B,Just15}, at least for nuclei with mass number $A>90$. On the other hand, under the hypothesis that all or most of the r-process material in the Universe is produced in BNS mergers,  the observed r-process abundances can be used to infer the merger rate \citep{2016MNRAS.455...17V}. The result of this estimate is indeed consistent with the merger rate measured directly from the two BNS mergers discovered in GW, which further supports the idea that most of the r-process elements were formed in these astrophysical sites.

However, a detailed comparison of the observed abundances with model predictions reveals a more complex picture.  Population synthesis models generally predict long time delays between star formation episodes and BNS mergers, on the order of hundreds of Myr and  up to several Gyr \citep{2018arXiv181210065B,2017AcA....67...37C,2019MNRAS.482.2234G}. These long time delays come about because the initial separations of the newly formed neutron stars can be quite large, and energy loss  to GW emission (and therefore orbital decay) is slow. This characteristic property of BNS mergers is difficult to reconcile with the obsevations of  ultra-metal-poor but r-process-rich stars in the MW halo and the Reticulum II and Tucana III ultra-faint dwarf galaxies \citep{2016ApJ...830...93J,2017ApJ...838...44H}. This early enrichment seems to suggest a common origin of iron-group and r-process elements in these stars, as pointed out in several studies \citep{2004A&A...416..997A, 2015MNRAS.447..140V,2017MNRAS.471.2088S, 2018MNRAS.477.1206N, 2018ApJ...855...99C,2019ApJ...876...28S,2020MNRAS.494.4867V}, although note that some earlier models managed to reproduce observations with BNS mergers as the only source of r-process elements \citep{2015ApJ...807..115S,2016ApJ...830...76K,2017MNRAS.466.2474H}.  Moreover, \citet{2019MNRAS.490..296B} show that the mass fraction of r-process elements that remains in dwarf galaxies can be as low as $10^{-3}$, further reducing their abundance in low-metallicity systems \citep[although see][]{2018MNRAS.478.1994B}. 
The role of BNS mergers in high-metallicity environments is also a matter of debate. \citet{2019ApJ...875..106C} argue that the decreasing trend of r-process elements like Europium relative to iron abundance observed in MW disk stars is inconsistent with the large time delays expected in the BNS scenario. On the other hand, \citet{2020arXiv200704442B} argue that this problem can be resolved by including the effect of natal kicks of the NSs. 

It is important to stress that in order to explain both the mean observed r-process abundance and its scatter relative to the iron-group element abundances, chemical evolution models need to include also a description of the dispersion of r-process elements in the interstellar medium (ISM), which can itself depend on the properties (rate, spatial distribution, explosion energy etc.) of r-process producing sites \citep{2018MNRAS.477.1206N,2019ApJ...876...28S,2019ApJ...871..247B,2018MNRAS.478.1994B,2019ApJ...878..125T}.

In this paper, we study the contribution of BNS mergers to the total r-process element abundances in a MW-like galaxy using a semi-analytic galaxy evolution model that includes a turbulent mixing scheme. We focus here on Europium (Eu), since it is produced almost exclusively through the r-process, and is readily observable in the spectra of stellar atmospheres. Our model is based on the work described in \citet{2016MNRAS.455...17V}, where many of the uncertainties are studied in detail. In the present paper, we extend the model to a full semi-analytic description of the growth of a MW-like galaxy, starting from small building blocks at early epochs. We then focus on the comparison between instantaneous and turbulent mixing, as well as the impact of the time-delay distribution of BNS mergers.

The structure of the paper is as follows. In Section \ref{sec:model} we discuss the main components of our model and the resulting BNS merger rates. We present the case of instantaneous mixing in Section \ref{sec:Eu}, where we focus on the differences between various time delay distributions.  Results for turbulent mixing, where both the mean abundances and the observed scatter are reproduced, are discussed in Section \ref{sec:inhom}. The sensitivity  to the model parameters is studied in Section \ref{sec:parstudy}. We conclude in Section \ref{sec:discussion}.

\section{Model description}
\label{sec:model}
\subsection{Merger trees of dark matter halos}
\label{sec:mt}

We use the modified GALFORM code \citep{2000MNRAS.319..168C,2008MNRAS.383..557P} to construct merger trees of dark matter (DM) halos that host the galaxies. This algorithm implements the excursion set formalism, a key aspect
of which is the conditional mass function: the fraction of mass $f(M_1|M_2)$ from halos of mass $M_2$ at redshift $z_2$ that
consisted of smaller halos of mass $M_1$ at an earlier redshift $z_1$. The conditional mass function is given by (Lacey \& Cole 1993):
\begin{equation}
\begin{split}
	& f(M_1|M_2)\dd\ln M_1 = \\
	& \sqrt{\frac{2}{\pi}}\frac{\sigma_1^2(\delta_1-
\delta_2)}{(\sigma_1^2-\sigma_2^2)^{3/2}}\exp\left(-\frac{1}{2}\frac{(\delta_1-
\delta_2)^2}{(\sigma_1^2-\sigma_2^2)}\right) \frac{\dd\ln\sigma_1}{\dd\ln M_1}{\dd\ln M_1} , 
\end{split}
\label{eq:cond_mf}
\end{equation}
where $\sigma_i=\sigma^2(M_i)$ is the variance of the linear 
perturbation field smoothed on scale $M_i$, and $\delta_i$ is the 
critical density for spherical collapse at redshift $z_i$. This conditional mass function is used 
to calculate the mean number of progenitors of mass $M_1$ at redshift 
$z_1+dz_1$ of a halo of mass $M_2$ at $z_2=z_1$:
\begin{equation}
	\frac{\dd N}{\dd M_1}=\frac{1}{M_1}\left(\frac{\dd f}{\dd z_1}\right)_{z_1=
z_2}\frac{M_2}{M_1}{\dd z_1} .
\label{eq:mean_numb}
\end{equation}
In the modified GALFORM algorithm the  following substitution is made:
\begin{equation}
	\frac{\dd N}{\dd M_1} \rightarrow \frac{\dd N}{\dd M_1}G(\sigma_1/\sigma_2,
\delta_2/\sigma_2) , 
\label{eq:G_function}
\end{equation}
where $G$ is referred to as a perturbing function, to be calibrated 
by comparison with N-body simulations. Parkinson et al. (2008) showed 
that by fitting the outcome of the algorithm to the results of the Millennium Simulation 
they obtained halo abundances which are consistent with the
Sheth-Tormen mass function \citep{2002MNRAS.329...61S}. Thus, the perturbing function
$G$ expresses the uncertainty in the choice of the correct mass function. In
this work we use the following parametrization of the perturbing function:
$G=G_0\left(\sigma_1/\sigma_2\right)^{\gamma_1}
\left(\delta_2/\sigma_2\right)^{\gamma_2}$, with the parameters $G_0=0.57$,
$\gamma_1=0.38$, $\gamma_2=-0.01$ taken from \citet{2008MNRAS.383..557P}.

By specifying the required mass resolution $M_\mathrm{res}$ we can now calculate the mean number of progenitors with masses $M_1$ in the range $M_{\rm res}<M_1<M_2/2$:
\begin{equation}
 P = \int_{M_{\rm res}}^{M_2/2}\frac{\dd N}{\dd M_1} {\dd M_1}
\end{equation}
as well as the fraction of mass of the final halo in unresolved objects:
\begin{equation}
 F = \int_{0}^{M_\mathrm{res}}\frac{\dd N}{\dd M_1} \frac{M_1}{M_2}{\dd M_1}\:.
 \label{eq:Funres}
\end{equation}
Starting with the specified mass and redshift, the algorithm proceeds 
back in time. The redshift step $\mathrm{d}z_1$ is chosen so that $P\ll 1$ to ensure that the halo is unlikely
to have more than two progenitors at the earlier redshift. The next step is to generate a uniform random number $R\subset(0,1)$. If $R>P$, no merger has occurred in this timestep and the halo mass is reduced to $M_2(1-F)$ (that is, the halo has grown by accretion of this unresolved mass rather than by mergers). If $R\leq P$, a merger has occurred, and a random value of the progenitor mass $M_1$ is generated using the distribution in Eq. (\ref{eq:mean_numb}). The merging masses in this case are $M_1$ and $M_2(1-F)-M_1$ (there is still accretion). This process is repeated until the progenitor mass at some node reaches the resolution mass (and so cannot be further subdivided) or until the limiting redshift (typically $z=15$) is reached.

Halo masses are stored only at specified redshifts. The final information is therefore contained in a list of halos and the redshifts of their mergers. Specifically, at each node we obtain the masses of the merging halos $M_1$ and $M_1'$ as well as the mass accreted onto them and the final big halo $M_2$:
\begin{equation}
 \Delta M_{\rm unres} = M_2-M_1-M_1'~.
 \label{eq:Munres}
\end{equation}
In the remainder of the paper we build merger trees for a DM halo with mass $10^{12}M_{\odot}$ at $z=0$ in order to match a MW-like structure \citep{1998gaas.book.....B,2016ARA&A..54..529B} and a resolution mass of $10^9M_{\odot}$. We checked that the results are converged at this resolution. The effect of the total DM halo mass is further discussed in Section \ref{sec:galmass}.

\subsection{Gaseous phases and star formation}
\label{sec:evolution}

\begin{figure}
\centering
\includegraphics[scale=0.4, angle=270]{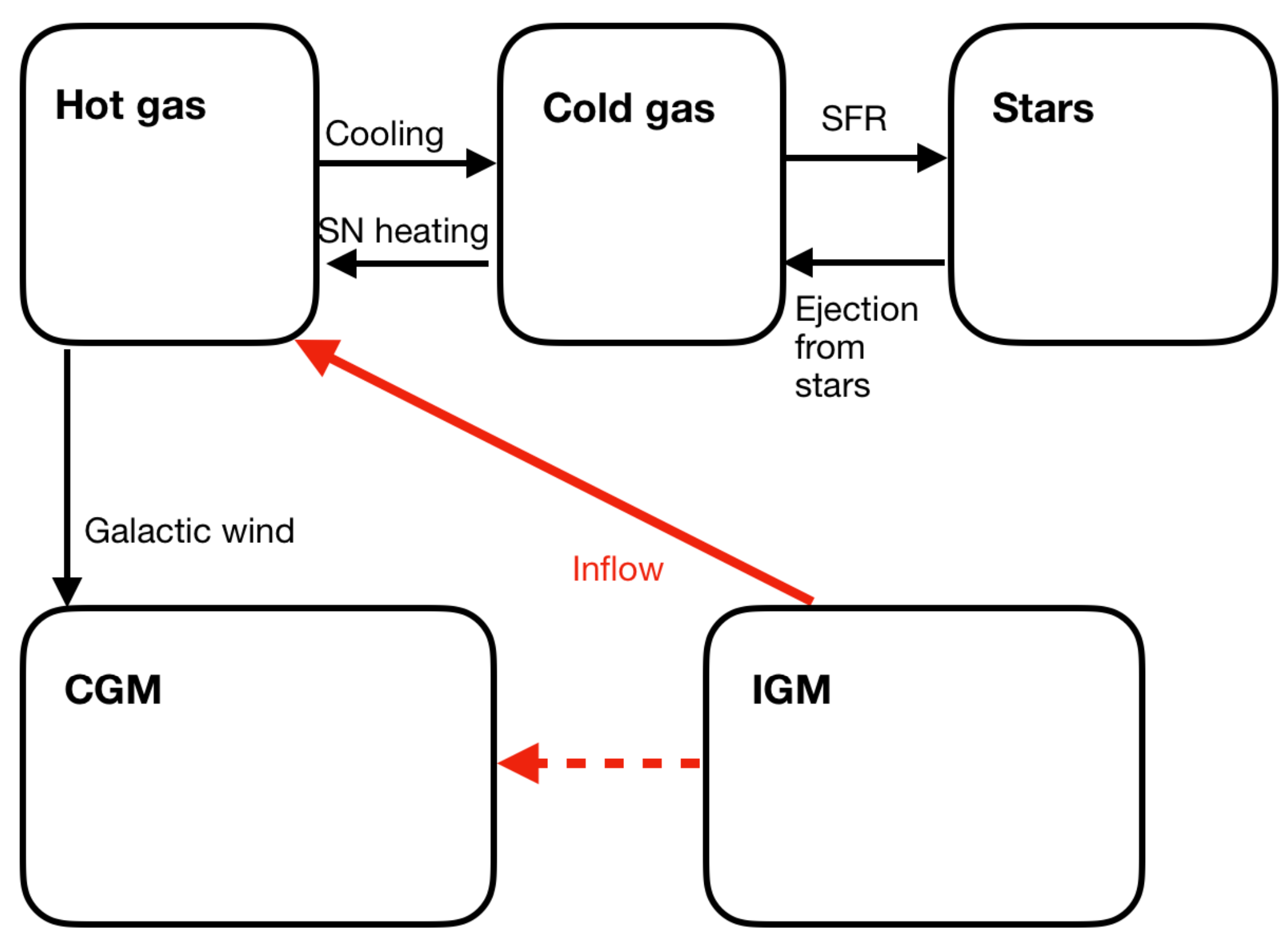}
\caption{Evolution of the galaxy along a merger tree branch. The infall from IGM into structures is given by the merger tree (shown by the red arrows). All other processes are calculated as described in the text. The CGM acts as a reservoir for gas outflows from the galaxy.
}
\label{fig:gal_evo}
\end{figure}

As in other semi-analytic galaxy evolution models \citep[e.g.][]{2000MNRAS.319..168C,2008MNRAS.391..481S,2010MNRAS.405.1573B,2012MNRAS.423.2533B,2013MNRAS.435.3500Y}, the quiescent evolution of the galaxy is obtained by solving a set of evolution equations on the branch (i.e. between mergers). The galaxy and its immediate surroundings are modelled as shown schematically in Fig. \ref{fig:gal_evo}. The galaxy contains $4$ reservoirs which correspond to the different phases: the circumgalactic medium (CGM), hot gas that occupies the halo, cold gas and stars whose (time-dependent) masses are $M_{\rm CGM}$, $M_{\rm hot},M_{\rm cold}$ and $M_{*}$, respectively. The baryonic phases occupy a DM halo with mass $M_\mathrm{DM}$. Furthermore, each baryonic phase $a$ is characterized by a metallicity $Z_a$. The CGM acts as a reservoir that contains gas outflows. Its mass at the beginning of each timestep is defined as the mass contained in an annulus between the virial radius $R_{\rm vir}$ of the halo and $f_{\rm CGM}\times R_{\rm vir}$, where $f_{\rm CGM}$ can be varied. The density profile outside of the halo is assumed to be an extrapolation of the one inside (i.e. a Navarro–Frenk–White \citep[NFW;][]{1996ApJ...462..563N} for DM, see below).

The gas flowing into the galaxy is equal to the sum of all the unresolved halos that 'merged' with the galaxy during the timestep. For a timestep $\Delta t_{\rm step}$ (determined by the merger tree resolution) the inflow rate is given by:
\begin{equation}
 \dot{M}_{\rm inflow}=f_b\frac{\Delta M_{\rm unres}}{\Delta t_{\rm step}}~,
\end{equation}
where $f_b$ is the baryon fraction and $\Delta M_{\rm unres}$  is given by Eq. (\ref{eq:Munres}). We assume that a fraction $\alpha_{\rm in}$ flows directly from the intergalactic medium (IGM) into the galaxy (for example via cold flows), as shown by the solid red arrow in Fig. \ref{fig:gal_evo}, and a fraction $1-\alpha_{\rm in}$ flows into the CGM, as shown by the dashed red arrow.  

The DM has an NFW profile with concentration $c$ \citep{2014MNRAS.441.3359D}:
\begin{equation}
 \rho_\mathrm{DM}(r)=\rho_s\left(\frac{r}{r_s} \right)^{-1}\left(1+\frac{r}{r_s} \right)~,
\end{equation}
where $r_s=R_{\rm vir}/c$, $R_{\rm vir}$ is the virial radius, $\rho_s=M_{\rm vir}/4\pi r_s^3 f(c)$ and $f(c)=\ln{(1+c)}-c/(1+c)$. We take $c=10$ for all the halos in the tree.

The gas that flows into the galaxy is assumed to be immediately shock-heated to the virial temperature $T_{\rm vir}=GMm_p/k_B R_{\rm vir}$ and to have an isothermal profile:
\begin{equation}
 \rho_\mathrm{hot}(r)=\rho_0\exp\left[-\frac{27}{2}\beta \left(1-\frac{\ln{(1+r/r_s)}}{r/r_s} \right) \right]
 \label{eq:rho_hot}
\end{equation}
with 
\begin{equation}
 \beta = \frac{8\pi\mu m_p G \rho_s r_s^2}{27k_B T_{\rm vir}}
\end{equation}
where $\mu$ is the mean molecular weight of the gas, $k_B$ is the Boltzmann constant, $m_p$ the proton mass and $\rho_0$ is a normalization given by the total gas mass in the halo.

Radiative cooling of the hot gas occurs on a timescale:
\begin{equation}
 t_{\rm cool} = \frac{3\rho_{\rm hot}k_BT_{\rm vir}}{2\mu m_p n_e^2 \Lambda(T_{\rm vir},Z_{\rm hot})}~,
\end{equation}
where $\Lambda$ is the cooling function \citep{1993ApJS...88..253S} and $n_e$ is the electron density. Then the rate of ``flow'' of  gas from the hot phase to the cold one is:
\begin{equation}
 \dot{M}_{\rm cooling} = 4\pi \int_0^{R}\frac{r^2\rho_{\rm hot}}{t_{\rm cool}}{\rm d}r~.
 \label{eq:coolRate}
\end{equation}

To model the rate at which stars form from cold gas we use a power-law scaling with a cut-off at high gas masses \citep{1998ApJ...498..541K}:
\begin{equation}
 \psi = A_{\rm SFR}\left(\frac{M_{\rm cold}}{M_{\rm 0, SFR}}\right)^\gamma e^{-M_{\rm cold}/M_{\rm q}}
 \label{eq:SFR_law}
\end{equation}
where $A_{\rm SFR}$, $\alpha$ and $M_{\rm q}$ are to be specified. The cut-off at $M_{\rm q}$ is meant to mimic the decline in the star formation rate (SFR) of the MW towards the present time \citep{2015A&A...578A..87S}. This value for the damping mass is motivated by observational and theoretical results \citep[e.g.][]{2013MNRAS.434..209B,2015MNRAS.446..521S}; we further discuss this parametrization in Section~\ref{sec:parstudy}.

The initial mass function (IMF) of newly-formed stars $\phi(m)$ is given by a Salpeter power-law $\phi(m)\propto m^{-p}$ with $p=2.35$ \citep{1955ApJ...121..161S} and normalized as follows:
\begin{equation}
 \int_{m_{\rm inf}}^{m_{\rm sup}} m \phi(m)\,{\rm d}m =1\:.
 \label{eq:IMFnorm}
\end{equation}
Stars return some of their mass into the cold phase in the form of winds, SN ejecta and planetary nebulae. This ``recycled'' mass flow is given by:
\begin{equation}
 e_{\rm rec} = \int_{m_{\rm inf}}^{m_{\rm sup}} \mathrm{d}m\, \phi(m)\psi(t-\tau(m,Z))(m-m_\mathrm{rem})
\end{equation}
where $Z$ is the metallicity, $m$ is the stellar mass, $m_\mathrm{rem}(m,Z)$ is the remnant mass, which we take from \citet{2012ApJ...749...91F} and $\tau(m,Z)$ is the stellar lifetime, taken from \citet{1989A&A...210..155M} and \citet{2002A&A...382...28S}.

\begin{table}
\begin{tabular}{c|c|c}
\centering
Name & Description & Value \\
\hline
\hline
$\alpha_{\rm in}$ & IGM fraction falling directly into galaxy & $0.8$ \\
$A_{\rm SFR}$     & SFR efficiency $[M_{\odot}/\mathrm{yr}]$    & $0.03$ \\
$M_{\rm 0,SFR}$ & Pivot mass  for SFR $[M_{\odot}]$  & $10^9$ \\
$M_{\rm q}$                  & Cut-off mass for SFR $[M_{\odot}]$      & $8\cdot 10^9$ \\
$\gamma$           & SFR slope: $\mathrm{d}\ln{\rm SFR}/\mathrm{d}\ln M_{\rm cold}$       & $2.0$ \\
$\epsilon_{\rm heating}$ & Heating efficiency                                               & $0.01$ \\
$m_{\rm inf}$      & Lower limit for stellar masses $[M_{\odot}]$                & $0.1$ \\
$m_{\rm sup}$      & Upper limit for stellar masses $[M_{\odot}]$                & $100$ \\
$f_{\rm ex}$        &  Fraction of reheated gas as galactic winds   & $0.05$ \\
$\alpha_{\rm WD}$ & Fraction of WD that merge as SNIa & $0.09$ \\
$t_{\rm min,WD}$ & Minimal time delay of SNIa [Myr] & $100$ \\
$m_{\rm Fe}$ & Fe mass produced in each SNIa $[M_{\odot}]$ & $0.5$ \\
$\alpha_{\rm BNS}$ & Fraction of NS in merging binaries & $4\cdot 10^{-3}$ \\
$m_{\rm Eu}$ & Eu mass produced in each merger $[M_{\odot}]$  & $10^{-4}$ \\
\hline

\end{tabular}
\caption{Parameters used in the galaxy evolution model (see text for details).
}
\label{tab:model_params}
\end{table}

In addition, SN-powered winds heat the cold gas at a rate
\begin{equation}
 \dot{M}_{\rm heating}=\frac{2\epsilon_{\rm heating}}{v_{\rm rot}^2} \int_{8\:M_{\odot}}^{m_f} \mathrm{d}m\, \phi(m)\psi(t-\tau(m,Z))E_{\rm kin}(m)
\end{equation}
where $E_{\rm kin}(m)$ is the energy released by the explosion of a star of mass $m$ taken from \citet{1995ApJS..101..181W}, $v_{\rm rot}$ is the rotation velocity of the DM halo.
A fraction $f_{\rm ex}$ of this reheated gas is expelled out of the halo as galactic wind. 

To summarize, the flow of gas between the different phases is given by the following equations:
\begin{eqnarray}
 \dot{M}_{\rm CGM} = \left(1-\alpha_{\rm in}\right)\dot{M}_{\rm inflow} + f_{\rm ex}\dot{M}_{\rm heating}\\
 \dot{M}_{\rm hot} = \alpha_{\rm in}\dot{M}_{\rm inflow}-\dot{M}_{\rm cooling} + \dot{M}_{\rm heating}(1-f_{\rm ex})\\
 \dot{M}_{\rm cold} = \dot{M}_{\rm cooling} - \dot{M}_{\rm heating} - \psi + e_{\rm rec} \\
 \dot{M}_* = \psi - e_{\rm rec}
\end{eqnarray}
Note that 
\begin{equation}
\dot{M}_{\rm CGM}+\dot{M}_{\rm hot}+\dot{M}_{\rm cold} + \dot{M}_* = \dot{M}_{\rm inflow}\, .
\end{equation}
Model parameters used in our calculation are listed in Table~\ref{tab:model_params}. We explore variations in DM halo mass (and stellar mass) as well as mass and time resolution in Section \ref{sec:parstudy}.

With the parameters summarized in Table \ref{tab:model_params}, our reference model for a $10^{12}M_{\odot}$ DM halo results in $M_*=10^{10}M_{\odot}$ in stars at $z=0$, close to the bulge mass in the MW estimated from observations \citep{2015ApJ...806...96L}. In addition, the SFR at $z=0$ is $1.22M_{\odot}/{\rm yr}$ in our model, compared with the observed value of $1.65M_{\odot}/{\rm yr}$ in the MW. The stellar mass in our model is thus lower than the total stellar mass in the MW (around $6\times 10^{10}M_{\odot}$). We further discuss this 
difference in Section \ref{sec:galmass}, where we show that our main conclusions are insensitive to the total stellar mass at $z=0$. We note here that the metallicity distribution function (MDF) in our model provides a reasonably good fit to the global MDF of the MW (see Section~\ref{sec:galmass}).

\subsection{Formation and merger rates of binary neutron stars}
\label{sec:bns}

The evolution of individual binary systems is best described with population synthesis models that follow the evolution of binary massive stars \citep[e.g.][]{2015ApJ...801...32A,2015ApJ...814...58D,2018MNRAS.474.2937C,2018arXiv181210065B,2019MNRAS.482.5012C,2019MNRAS.487.1675A,2019MNRAS.482.2234G}. The results of these models are then implemented in galaxy evolution models to obtain the merger rate. Since this calculation involves many different spatial and temporal scales,  each model necessarily involves various simplifications in the description of BNS formation or galaxy evolution. The outcome of the most recent models is generally in agreement with the local rate at $z=0$ of $1090_{-800}^{+1720}$ Gpc$^{-3}$yr$^{-1}$ inferred from the gravitational-wave observations of two BNS mergers GW170817 and GW190425  \citep{2017PhRvL.119p1101A,2020ApJ...892L...3A}. Note that BNS can also form in globular clusters \citep{2020ApJ...888L..10Y} or nuclear clusters \citep{2019MNRAS.488...47F}, however the formation rates in these channels are expected to be several orders of magnitude below the observed value.

Another observational constraint can be inferred from the observations of the Galactic double binary pulsars \citep{2015MNRAS.448..928K,2020RNAAS...4...22P}. The latest estimate of $R_g=37^{+24}_{-11}$ Myr$^{-1}$ from \citet{2020RNAAS...4...22P} is consistent with the local rate measured from GW observations, although note that the translation of this Galactic rate into the local rate requires an understanding of the dependence of the BNS merger rate on galactic properties.

\begin{figure}
\centering
\includegraphics[width=0.45\textwidth]{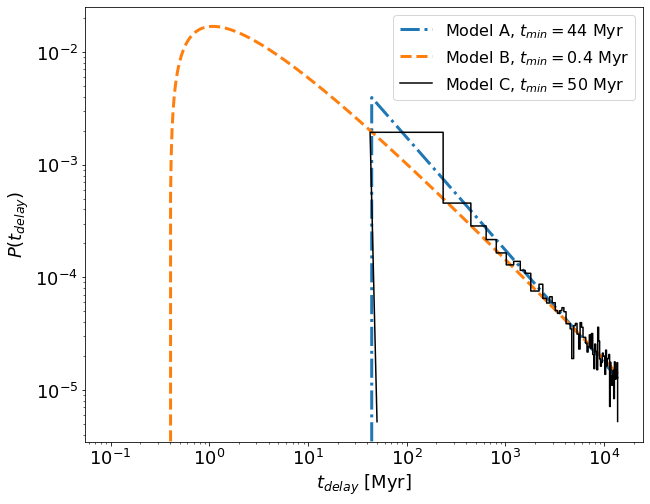}
\caption{Time delay distributions between formation and merger of the BNS. Models A, B are analytical, while Model C uses the data from Fig. 5 in \citet{2018MNRAS.474.2937C}. The value of $t_{\rm min}$ for Models A and B were chosen to give the same mean time delay as Model C with $t_{\rm min}=50$~Myr, {\it i.e.} $\langle t_{\rm delay} \rangle = 2.37$~Gyr.
}
\label{fig:time-delays}
\end{figure}

In this work we use a phenomenological approach to describe BNS formation rate. At each time step of the calculation, and for a given stellar mass at birth $m$ and metallicity $Z$ (chemical evolution is discussed below) we calculate the remnant mass $m_{\rm rem}(m,Z)$ using the results of \citet{2012ApJ...749...91F}. Namely, the remnant mass is calculated separately for stars with initial masses below $11M_{\odot}$ and $11M_{\odot}<m<30M_{\odot}$ (we assume that all stars with initial masses above $30M_{\odot}$ collapse to a BH). The lower mass limit for NS formation is taken to be:
\begin{eqnarray}
 m_{\rm NS}^{\rm lower}=9M_{\odot}+0.9\log_{10}(Z_{\rm met}/Z_{\odot})M_{\odot}
\end{eqnarray}
for metallicities of $Z_{\rm met} > 10^{-3}Z_{\odot}$ and $m_{NS}^{\rm lower}=6.3M_{\odot}$ otherwise. Thus, all stars in the initial mass range of $m_{\rm NS}^{\rm lower}<m<11M_{\odot}$ form a NS. The mass remnant of stars in the initial mass range of $11M_{\odot}<m<30M_{\odot}$ is given (in the delayed explosion scenario) by:
\begin{equation}
    m_{\rm rem}=1.1+0.2e^{(m-11)/4}-(2+Z_{\rm met})e^{0.4(m-26)}
\end{equation}
where both $m$ and $m_{\rm rem}$ are in units of the solar mass. We then take any remnant with $m_{\rm rem}<3M_{\odot}$ to be a NS.
Below $m_{NS}^{\rm lower}$ all stars become white dwarfs.

Then the formation rate of NSs is given by 
\begin{equation}
 R_{\rm NS}(t)=\int_{m_{\rm inf}}^{m_{\rm sup}} \epsilon_\mathrm{NS}(m,Z)\psi(t-\tau(m,Z)) \phi(m)\, \rm{d} m  ~,
\end{equation}
where $\epsilon_\mathrm{NS}(m,Z)$ is the probability to form a NS, taken here to be $1$ if the identity of the remnant is a NS according to the fitting function in \citet{2012ApJ...749...91F} and $0$ otherwise.
Having obtained the formation rate of NSs $R_{\rm NS}(t)$, we  assume that a fraction $\alpha_{\rm BNS}$ of these NSs are in binaries that merge within the age of the Universe after a delay time $t_{\rm delay}$. The merger rate at time $t$ is given by
\begin{equation}
 R_{\rm merger}(t)=\alpha_{\rm BNS}\int_{t_{\rm min}}^{t_{\rm max}} R_\mathrm{NS}(t-t_{\rm delay}) P(t_{\rm delay}) \,\mathrm{d}t_{\rm delay}
 \label{eq:Rmerger}
\end{equation}
where $t_{\rm min}$ and $t_{\rm max}$ are specified below. Note that the factor $\alpha_{\rm BNS}$ accounts for the formation of binaries.

In the following we consider three models for the delay time distribution $P(t_{\rm delay})$ (see Fig. \ref{fig:time-delays}). Our Model A corresponds to the standard $1/t$ normalized distribution, {\it i.e.}
\begin{equation}
 P(t_{\rm delay}) = A t_{\rm delay}^{-1},\: t_{\rm delay}\in (t_{\rm min}, t_{\rm max})
\end{equation}
where $A =1/ \log (t_{\rm max}/t_{\rm min})$. This distribution is characterized by a mean delay time given by $\langle t_{\rm delay} \rangle=(t_{\rm max}-t_{\rm min})/ \log (t_{\rm max}/t_{\rm min})$.
The choice $t_{\rm max}=t_H\simeq 13.8$ Gyr is a natural cut-off, since these models are not evolved past the current age of the Universe. Note that an extrapolation of this distribution to even longer delay times can be counterbalanced by simply decreasing the total merging fraction $\alpha_\mathrm{BNS}$. This functional form is a generic choice in population studies, as it is often obtained in population synthesis models \citep[e.g.][]{2012ApJ...759...52D}.

Due to the crucial character of the delay time distribution at early times, {\it i.e.} in the vicinity of $t_{\rm min}$, we also consider here another type of distributions which allows for a more detailed account of BNS mergers with small delay times without the artificial cut-off at $t=t_{\rm min}$. This Model B is described by the functional
\begin{equation}
 P(t_{\rm delay}) = B \frac{\ln{\left(t_{\rm delay}/t_{\rm min}\right)}}{t_{\rm delay}},\: t_{\rm delay} \in (t_{\rm min}, t_{\rm max})
\end{equation}
with a normalization constant $B = 2/(\ln (t_{\rm max}/t_{\rm min}))^2$ leading to a mean time delay of $\langle t_{\rm delay}\rangle= B t_{\rm max}\left(\ln(t_{\rm max}/t_{\rm min}) + t_{\rm min}/t_{\rm max} -1 \right)$. This functional form provides a good fit to the recent population synthesis model of  \citet{2018MNRAS.474.2937C}, at least for $t_{\rm delay}\ga 100$~Myr, and has the particularity to include a soft cut-off at low $t_{\rm delay}$. This specific $\ln(t)/t$ behaviour of the distribution allows us to consider a non-negligible fraction of BNS mergers at early times, as discussed below without affecting the $1/t$ law found by population synthesis models (see Fig.~\ref{fig:time-delays}).

Finally, we also implemented a time delay distribution in tabulated form, taken directly from Fig. 5 in the recent detailed population synthesis model of \citet{2018MNRAS.474.2937C}, which we refer to as Model C. Note that the actual minimal delay time remains uncertain in such an approach and may be an artifact of the binning (linear in time) in \citet{2018MNRAS.474.2937C}. We adopt for this model an arbitrary value of $t_{\rm min}=50$~Myr. 

All three models are compared in Fig.~\ref{fig:time-delays} assuming the same value of $t_{\rm max}=13.8$~Gyr and a value of $t_{\rm min}$ tuned, so that the three distributions have the same mean delay time of $\langle t_{\rm delay} \rangle=2.37$~Gyr.  As seen in Fig.~\ref{fig:time-delays}, despite their similar mean time delay, these models can have a very different behaviour at early times, hence a different fraction of mergers with very short time delays. Such a fraction of systems with a time delay $t_{\rm delay} \le t^*$ is given by 
\begin{equation}
    F_{\rm delay}(<t^*)=\int_{t_{\rm min}}^{t^*} P(t_{\rm delay})~ dt_{\rm delay}. 
    \label{eq_frac}
\end{equation}
For example, for the parameters shown in Fig. \ref{fig:time-delays}, the fraction of systems with delay times below $t^*=100$~Myr amounts to $0.14$ for Model A, $0.27$ for Model B and $\sim 0.11$ for Model C. 
As shown in \citet{2016MNRAS.455...17V} (see in particular their Fig. 3), population synthesis models \citep{Belczynski02} predict that a significant fraction of BNS mergers takes place with short time delays, and more specifically that about 20 to 25\% of systems may merge with timescales lower than 100~Myr. Such a significant fraction of fast merging systems is properly described by our Model B shown in Fig.~\ref{fig:time-delays} but is underestimated by Models A and C.

At each time step $\Delta t$ we calculate the total mean number of BNS mergers $\langle N_m\rangle=R_{\rm merger} \Delta t$ using Eq. \ref{eq:Rmerger}. In the instantaneous mixing case (described in Section~\ref{sec:Eu}) we use $\langle N_m\rangle$ in our chemical evolution equations (see Section~\ref{sec:chem}). However, as BNS mergers are rare events, $\langle N_m\rangle$ can be small and even much smaller than $1$ in many time steps, especially at high redshifts for low-mass halos. This led us to develop a more realistic approach, namely the turbulent mixing model, where we draw the number of actual mergers from a Poisson distribution with mean $\langle N_m\rangle$As a consequence, for certain timesteps there are no mergers at all. Then, for each of these mergers, we draw its position in the galaxy, assuming a 3-dimensional Gaussian distribution with variance equal to the half-light radius $R_e$. The latter is computed as $R_e=0.02R_{\rm vir}$ \citep{2020MNRAS.492.1671Z}. 
As an output of this calculation, we obtain a list of BNS mergers, their host halos, and the time and spatial position of each merger. The results for our turbulent mixing case are described in Section~\ref{sec:inhom}.  

Finally, in both cases we assume that each BNS merger produces a (constant) mass $m_{\rm r}$ of r-process elements, and more specifically a Eu mass $m_{\rm Eu}=10^{-4}M_{\odot}$, as predicted by realistic BNS merger simulations of both the dynamical and wind ejecta \citep{Just15}. Note that the only real free parameter governing the evolution of Eu in our model is the product $\alpha_\mathrm{BNS}\,m_\mathrm{Eu}$We checked that 
the mean evolution is
insensitive to changes in $\alpha_\mathrm{BNS}$, $m_\mathrm{Eu}$ that keep their product constant and discuss in Section \ref{sec:inhom} the impact of $\alpha_\mathrm{BNS}$ on the dispersion in the turbulent mixing scenario.

\begin{figure}
\centering
\includegraphics[width=0.45\textwidth]{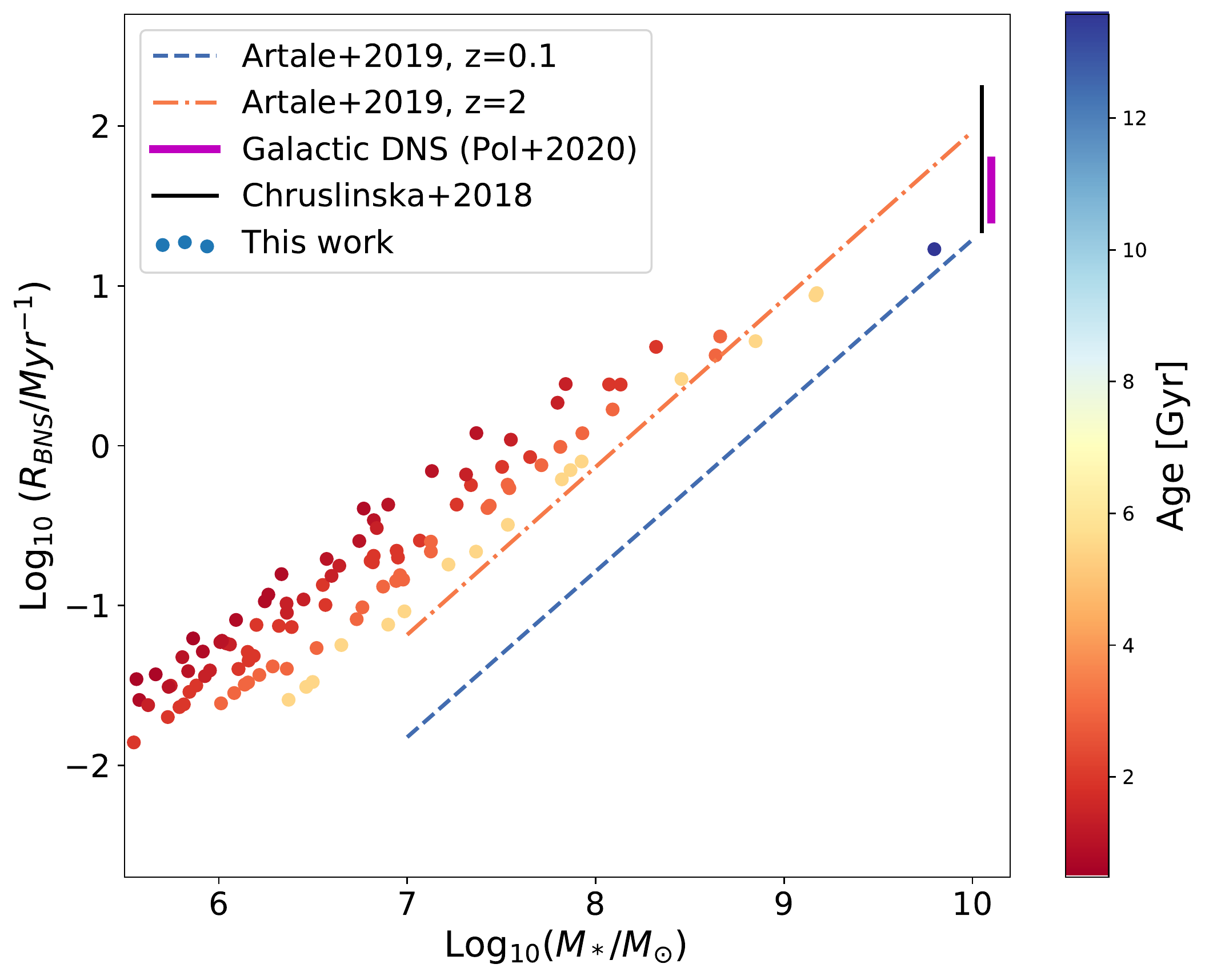}
\caption{BNS merger rates vs. stellar mass in the subhalos in the tree (colored dots) with time delay distribution according to Model A with $t_{min}=10$ Myr. The color code corresponds to the age of the subhalo (see colorbar). Vertical lines show the range of predictions in the models of  \citet{2018MNRAS.474.2937C} for a MW-like galaxy (plotted here for $M_*=10^{10}M_{\odot}$) and the galactic rates \citep{2020RNAAS...4...22P} based on the observation of the double pulsar. Also shown are the predictions computed from the model in \citet{2019MNRAS.487.1675A}  for $z=0.1$ and $z=2$. 
}
\label{fig:RNS}
\end{figure}

The merger rates in our MW-like galaxy model, for the three time delay distributions discussed above, are shown in Fig. \ref{fig:RNS}. We note that in our model, all the small halos that form at high redshifts merge with each other or fall onto the larger halo, and thus act as building blocks for the final, MW-like structure. Therefore, each point in Fig. \ref{fig:RNS} corresponds to a specific (sub-)halo and shows the instantaneous BNS merger rate at the cosmic age that corresponds to its formation (color-coded).

We also compare our results with the new constraints from Galactic double NSs \citep[][; magenta line]{2020RNAAS...4...22P} and two population synthesis models:  \citet{2018MNRAS.474.2937C} (black vertical line) and \citet{2019MNRAS.487.1675A} (color-coded lines).  For the comparison with the model in \citet{2019MNRAS.487.1675A}, which gives the merger rate as a function of stellar mass and redshift, we compute the merger rate for each subhalo in our model with stellar mass $M_*>10^7M_{\odot}$ (the limit given by  \citet{2019MNRAS.487.1675A}) and for two redshifts $z=0.1$ and 2 (color-coded). The result in \citet{2018MNRAS.474.2937C} is given only for $z=0$, and we plot it for $M_*=10^{10}M_{\odot}$. We note that our results are slightly below the value observed from Galactic double NSs at $z=0$, and the corresponding value in the \citet{2018MNRAS.474.2937C} model. On the other hand, our results are in general agreement with \citet{2019MNRAS.487.1675A}, given as a function of stellar mass. In particular, we observe that the merger rate for a given stellar mass slightly increases with redshift. We note that we made no particular attempt to fit the exact physical parameters of the MW as discussed in Section~\ref{sec:evolution} and \ref{sec:parstudy}.

\subsection{Chemical evolution}
\label{sec:chem}

First we outline the equations of chemical evolution for elements produced in stars, which follow those in \citet{2004ApJ...617..693D,2006ApJ...647..773D}, although note the differences due to different definitions and the introduction of gas heating and cooling.  

For each element $i$, the rate at which it is ejected from stars to the cold gas phase is:
\begin{equation}
 e_{i}(t) = \int_{m_{\rm inf}}^{m_{\rm sup}} \mathrm{d}m \phi(m)\psi(t-\tau(m,Z))m_i^{\rm ej}(m,Z)
\end{equation}
where $m_i^{\rm ej}(m,Z)$ is computed from stellar yieldstaken from \citet{1995ApJS..101..181W}. In addition, BNS mergers produce Eu with the rate: $e_{\rm Eu}(t)=m_{Eu}R_{\rm merger}(t)$ where $R_{\rm merger}(t)$ is given in eq. \ref{eq:Rmerger} and $m_{Eu}$ is the mass in Eu produced in each merger (see Table \ref{tab:model_params}). In the instantaneous mixing model we treat this contribution in the same way as the other species, whereas in the turbulent mixing case we follow the scheme described in detail below. We assume that Eu is produced only in BNS mergers.

We define by $X_{\rm i,hot}$, $X_{\rm i,cold}$, $X_{\rm i,CGM}$ and $X_{\rm i,IGM}$ the mass fraction of element $i$ in the cold, hot, CGM and intergalactic medium (IGM) phases, respectively. We assume that matter in-falling into the galaxy, $\dot{M}_{\rm inflow}$, does not contain metals. The evolution of the metal mass fractions is consequently given by
\begin{equation}
 \dot{X}_{\rm i,cold}(t)=\frac{1}{M_{\rm cold}}\left[\dot{M}_{\rm cooling}\left(X_{\rm i,hot}-X_{\rm i,cold} \right)-X_{\rm i,cold}e_{\rm rec}+e_i \right]
\end{equation}
\begin{equation}
\begin{split}
 \dot{X}_{\rm i,hot}(t)=\frac{1}{ M_{\rm hot}}\left[ \dot{M}_{\rm heating}\left( 1-f_{\rm ex}\right)\left(X_{\rm i,cold}-X_{\rm i,hot} \right)+\right. \\
    \left. \alpha_{\rm in}(X_{\rm i, IGM}-X_{\rm i,hot})\dot{M}_{\rm inflow} \right]
 \end{split}
\end{equation}
\begin{equation}
\begin{split}
 \dot{X}_{\rm i,CGM}(t)=\frac{1}{M_{\rm CGM}}\left[\dot{M}_{\rm heating}f_{\rm ex}\left(X_{\rm i,hot}-X_{\rm i,CGM} \right)+\right. \\ \left. (X_{\rm i,IGM}- X_{\rm i,CGM})(1-\alpha_{\rm in})\dot{M}_{\rm inflow}\right]
 \end{split}
\end{equation}
\vspace{5mm}

We calculate the evolution of iron abundance as in \citet{2016MNRAS.455...17V}; more  specifically, we consider two sources of iron, {\it i.e.} core-collapse SNe and type-Ia SNe. For the latter, we assume that a fraction $\alpha_{\rm WD}=0.09$ of the white dwarfs formed in our model ends up as type-Ia SNe, where the delay time between white dwarf formation and SNIa explosion, $t_{\rm d,wd}$, is distributed as $P(t_{\rm d,wd})\propto t_{\rm d,wd}^{-1}$ with $t_{\rm min, wd}=100$ Myr. We assume that each SNIa produces $0.5M_{\odot}$ in iron. A discussion on the impact of these assumptions can be found in \citet{2016MNRAS.455...17V}. With the model parameters summarized in Table \ref{tab:model_params} we obtain a rate of SNIa of $0.4$ per century at $z=0$, consistent with observations in the MW \citep{2014SAAS...37..145M}. The evolution of metallicity is further discussed below.

Our model allows us to reconstruct the chemical enrichment history of each galaxy. Indeed, the identity of each halo, its stellar mass and metallicity are saved at various time steps along the merger tree. Then by going backwards in the tree and tracing the times of past mergers we can reconstruct the time of formation of a certain stellar mass and subsequently its metallicity.

\subsubsection{Turbulent mixing of r-process elements}
\label{sec:inhomogen}

In order to model the production of r-process elements in BNS mergers and their subsequent diffusion in the surrounding ISM, we use the turbulent mixing scheme proposed by \citet{2015NatPh..11.1042H} and, recently, \citet{2020arXiv200301129B}. These authors propose an analytic description of the turbulent diffusion process that leads to the mixing of metals in the ISM. This mixing proceeds in two stages: first the newly synthesized metals expand with the blast wave until the end of the Sedov-Taylor phase. We neglect this short-lived phase. Then metals continue to diffuse due to turbulence in the ISM. The turbulent diffusion coefficient is given by \citep{2020arXiv200301129B}:
\begin{equation}
 D=\alpha c_s H=\alpha\left(\frac{c_s}{10 \rm{km/s}} \right)\left(\frac{H}{100\: \rm{pc}} \right) \rm{kpc^2 Gyr^{-1}} \:,
 \label{eq:coefD}
\end{equation}
where $c_s$ is the sound speed in the ISM, $H$ is the scale height and $\alpha$ is the efficiency parameter. Typical diffusion length at time $t$ is then given by $\sqrt{Dt}$. Since we do not describe the 3D shape of the galaxy, in the following we treat the diffusion coefficient $D$ directly as the free parameter of the model which we assume to be constant in all structures. Below we explore value of $D$ in the range of $0.001$-$0.01\, \mathrm{kpc
^2Gyr^{-1}}$. 

By including the turbulent diffusion effect, \citet{2020arXiv200301129B} obtained promising results for the scatter in the r-process element abundances observed in the MW, although for their main results they considered the case of negligible delay times between formation and merger of the BNS, constant SFR and focused on metallicity values of $[{\rm Fe/H}]>-1$. In the present work, we implement turbulent mixing in our full galaxy evolution model and show that it can explain the observed scatter across the entire metallicity range.

Our calculation proceeds as follows.
As described in Section~\ref{sec:bns}, at each time step $\Delta t$ of the calculation we obtain a list of BNS mergers (drawn from a Poisson distribution with mean $\langle N_m\rangle=R_{\rm merger} \Delta t$, see Eq. (\ref{eq:Rmerger})), their host galaxies, and the (3D) positions $r_i$ and times $t_i$ of all the events.
We assume that each of these mergers produces a fixed amount of Eu $m_{\rm Eu}$, which then disperses in the ISM. If at a later time a star forms in the vicinity of the merger event, it will be enriched by the Eu found in the ISM. If the star forms farther away and/or much later than the BNS merger, the abundance of Eu it will inherit will correspond to the mean value at its location. Taking into account all the contributions from BNS mergers in a given galaxy, we can express the Eu density at a given location, hereafter referred to as ``observer point'', $\vec{r}_{\rm obs}$ at time $t_{\rm obs}$ by \citep[see Eq.~2 in][]{2020arXiv200301129B}:
\begin{equation}
 \rho_{Eu}(\vec{r}_{\rm obs},t_{\rm obs})=\sum_{i} \frac{m_{\rm r}\exp\left(-\frac{|\vec{r}_{\rm obs}-r_i|^2}{ 4D\cdot(t_{\rm obs}-t_i)} \right)}{\left( 4\pi D\cdot(t_{\rm obs}-t_i)\right)^{3/2}}
 \label{eq:diffeq}
\end{equation}
where $D$ is the diffusion coefficient, $\vec{r}_i$ and $t_i$ are the positions and times of the BNS merger, respectively, and the sum runs over all the mergers in the given galaxy such that $t_{\rm obs} > t_i$. Note that we use a simplified version of the expression given in \citet{2020arXiv200301129B}, in particular we assume spherical diffusion. We take a number of observer points proportional to the stellar massformed along the branch, between mergers. We take a maximum of $500$ observer stars per halo (we verified that the results are converged in this limit). Since the calculation involves random draws for the number of mergers, their positions and times, each realization of the turbulent mixing scheme for a given merger tree provides a different evolution of Eu abundance, contrary to the homogeneous case, as we further discuss below. We note that we do not employ the turbulent mixing scheme for SNII or SNIa, but rather assume that the metals produced by all SNe are immediately mixed in the entire galaxy. This assumption is justified by the fact that these events are very frequent relative to BNS mergers.

\subsection{Galaxy mergers}
\label{sec:merger}

During a merger event $n$ smaller galaxies fall onto a larger galaxy. We assume that the stars and DM are simply added up, while the cold gas of the smaller galaxies is heated and added to the hot gas reservoir:
\begin{eqnarray}
 M_{\rm DM}=M_{\rm DM,host}+\Sigma_{j=1}^{n} M_{\rm DM,j}\\
 M_{\rm CGM}=M_{\rm CGM,host}+\Sigma_{j=1}^{n} M_{\rm CGM,j}\\
 M_{\rm hot}=M_{\rm hot,host}+\Sigma_{j=1}^{n} \left(M_{\rm hot,j}+M_{\rm cold,j}\right)\\
 M_{*}=M_{\rm *,host}+\Sigma_{j=1}^{n} M_{*,j}
\end{eqnarray}
In the turbulent mixing scheme for r-process elements abundance, we assume that, after each merger event, all the gas of the in-falling galaxies is completely mixed, and in particular the r-process elements are distributed evenly in the whole ISM. This mean abundance (immediately after the merger) is calculated under the instantaneous mixing approximation discussed above  and therefore takes into account mass loss due to galactic outflows.

\section{Europium abundance: instantaneous mixing}
\label{sec:Eu}
In this section we explore the Eu abundance as a representative r-process element.
As discussed above, we build merger trees for a halo of $10^{12}M_{\odot}$ which results in a stellar mass of $M_*=10^{10}M_{\odot}$ at $z=0$, and  compare our results with the compilation of Eu and Fe abundances from the SAGA database\footnote{http://sagadatabase.jp/}~\citep{2008PASJ...60.1159S} where we took all the stars in the MW and dwarf galaxies present in the database and removed duplicates. We assume throughout a constant BNS merger fraction of $\alpha_\mathrm{BNS}=4\times 10^{-3}$, $m_{\rm Eu}= 10^{-4}M_{\odot}$. Our chemical evolution model results in a metallicity of [Fe/H]$ \,\simeq -0.1$ at $z=0$, slightly higher than \citet{2016MNRAS.455...17V} but still lower than the value reached by disk stars, which we further discuss in the following Section. We also verified that the evolution of $\alpha$-elements in our model is consistent with observations in the entire metallicity range.

In this Section we assume an instantaneous mixing of Eu in the ISM and focus on the predictions obtained with different time delay distributions. Even in this instantaneous mixing model, some scatter in the [Eu/Fe] vs. [Fe/H] plane is expected, especially at early times, due to the fact that small halos can have very different chemical enrichment histories.

\begin{figure}
\centering
\includegraphics[width=0.47\textwidth]{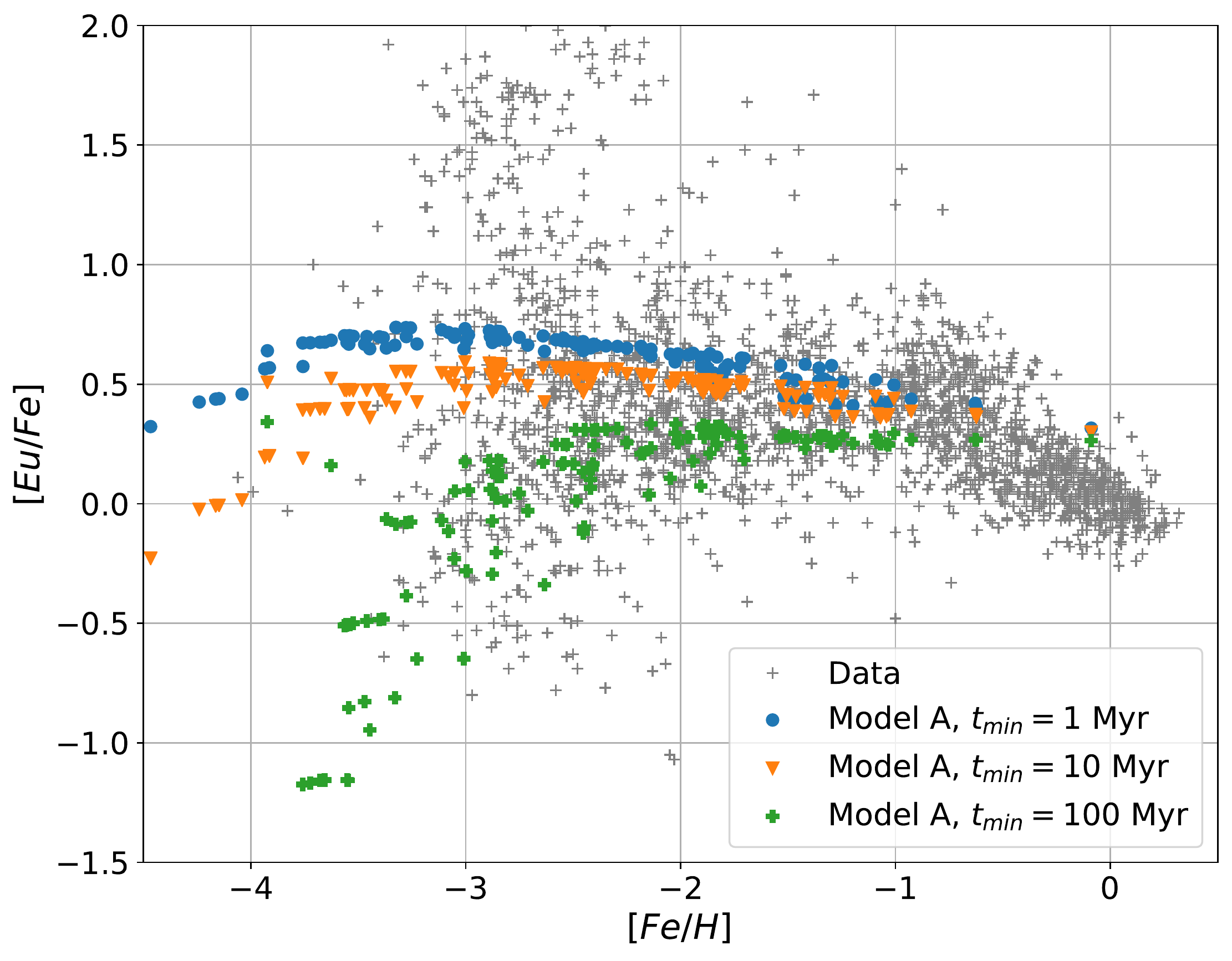}
\caption{Evolution of the Eu abundance in all the sub-halos in the tree with time delay distribution given by Model A: $P(t)\propto 1/t$. Blue circles are obtained with $t_{\rm min}=1$~Myr, orange triangles with $t_{\rm min}=10$~Myr and  green diamonds with $t_{\rm min}=100$~Myr. In all cases, $t_{\rm max}=13.8$~Gyr is adopted. Observations from the SAGA database are shown by grey crosses. 
}
\label{fig:one_over_t}
\end{figure}

Fig. \ref{fig:one_over_t} shows the results for Model A ({\it i.e.} the $t^{-1}$ time delay distribution) characterized by different values of $t_{\rm min}$. Each point corresponds to a single halo. The increase of the cut-off time $t_{\rm min}$ clearly sets the galactic Eu enrichment to higher metallicity, as already widely studied in many galactic chemical evolution models \citep{2014MNRAS.438.2177M,2019MNRAS.486.2896S,2019ApJ...875..106C}. In addition, we observe that longer time delays result in slightly larger dispersion in the [Eu/Fe] vs. [Fe/H] plane (the mean delay times for the three cases considered here are $1.4$, $1.9$, and $2.8$ Gyr for $t_{\rm min}=1$, $10$, and $100$ Myr, respectively) and that the scatter increases towards low metallicities.
The general shape of the evolution observed in Fig. \ref{fig:one_over_t} is qualitatively consistent with the conclusions of \citet{2020MNRAS.494.4867V}, who used an hydrodynamical simulation to model a MW-like galaxy and a phenomenological description of BNS formation similar to the one considered here, in particular a $t^{-1}$ time delay distribution in their fiducial model. 
However, we note that \citet{2020MNRAS.494.4867V} obtained a significantly larger scatter than the one shown in Fig. \ref{fig:one_over_t}, as discussed below.

\begin{figure}
\centering
\includegraphics[width=0.47\textwidth]{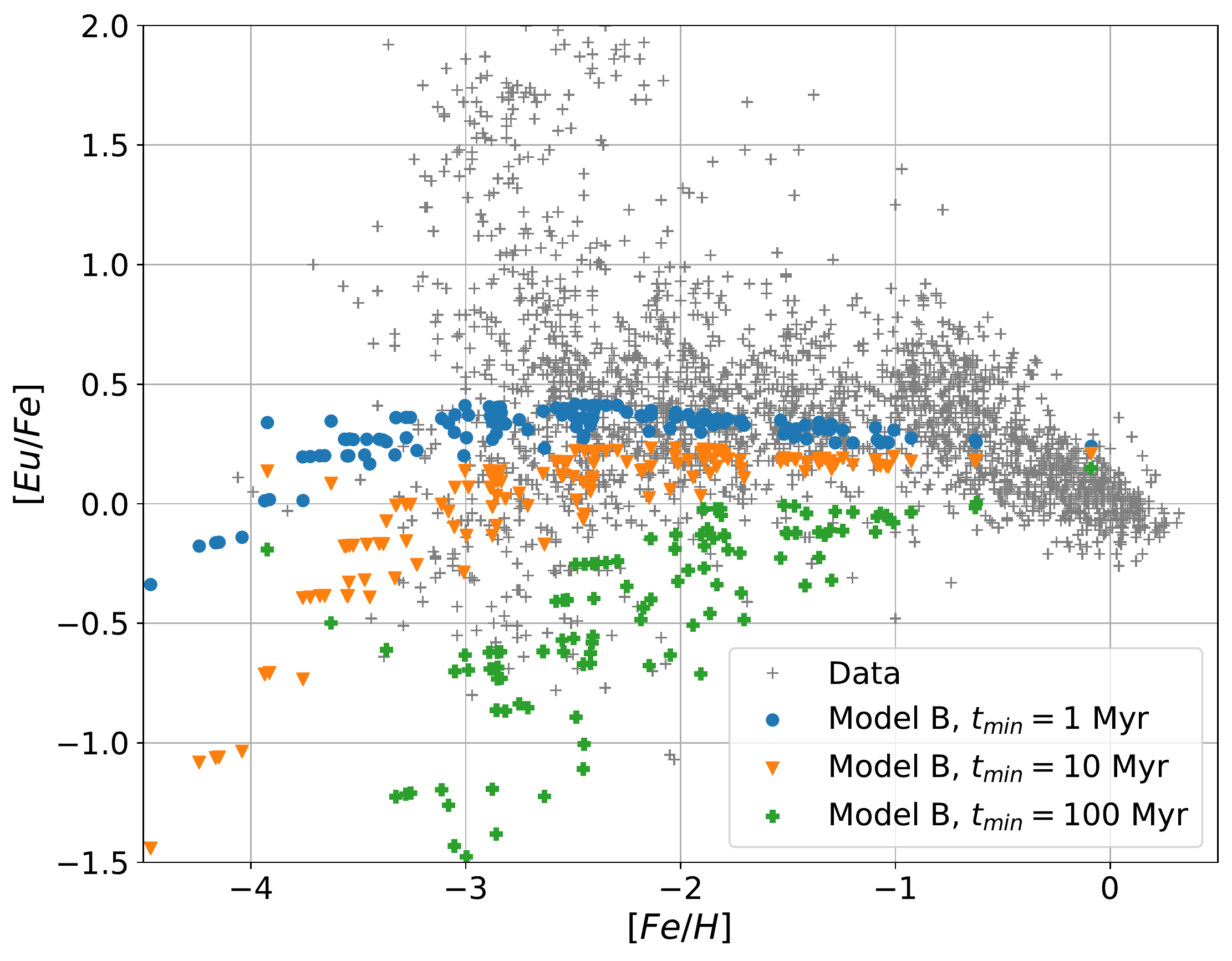}
\caption{Same as Fig. \ref{fig:one_over_t} but for Model B time delay distribution ($P(t)\propto \log(t/t_{\rm min})/t$). The results are qualitatively similar to those in Fig. \ref{fig:one_over_t}.}
\label{fig:logt_t}
\end{figure}

Next we explore other time delay distributions, not treated in previous works on r-process abundances. The results for Model B ($\log(t)/t$ time delay distribution) are shown in Fig. \ref{fig:logt_t} and are quite similar to those obtained with Model A. In this case, the dispersion at low metallicities and for large $t_{\rm min}$ is slightly larger than in the case of Model A, but the mean value is significantly lower.

\begin{figure}
\centering
\includegraphics[width=0.47\textwidth]{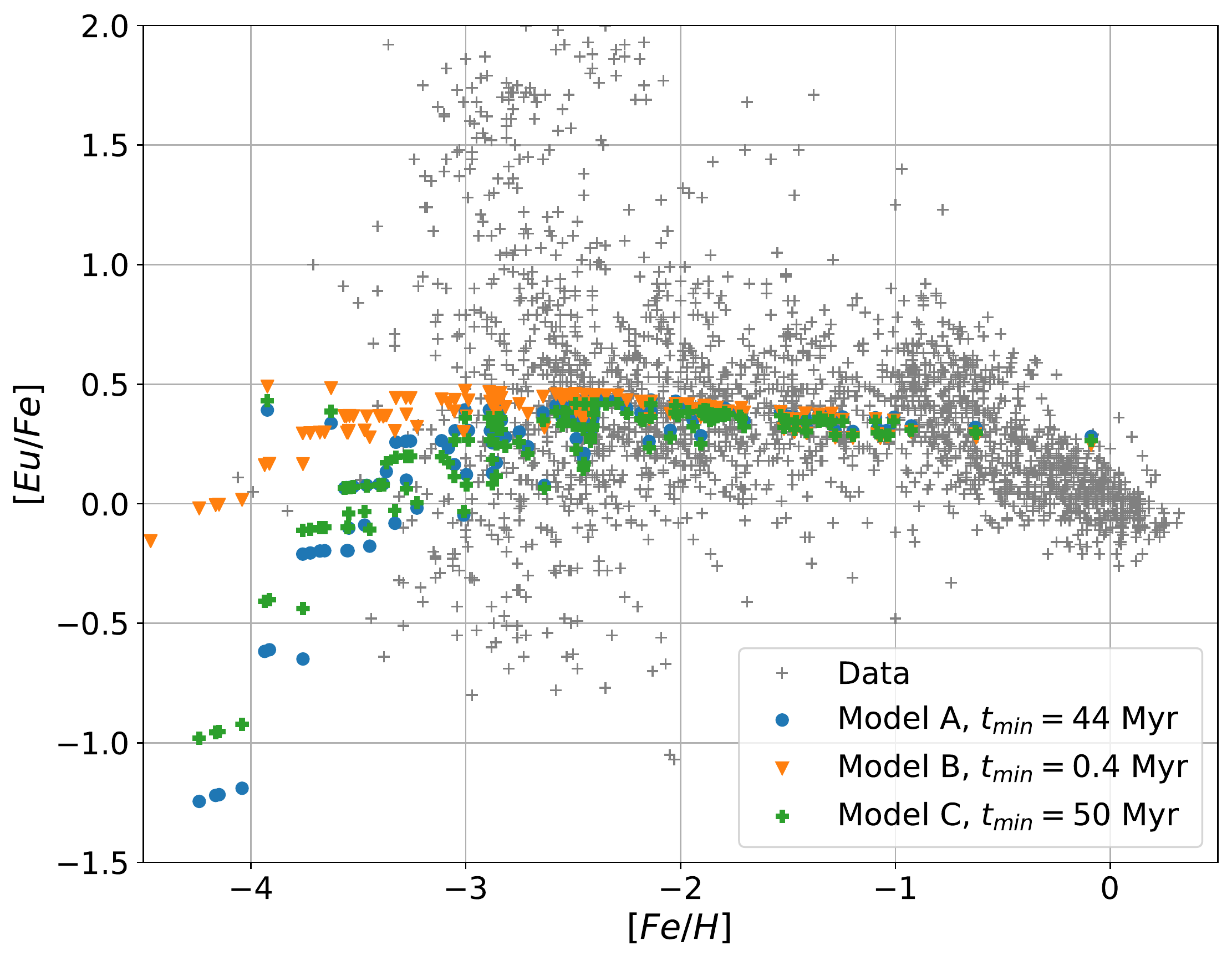}
\caption{Comparison of [Eu/Fe] versus [Fe/H] predictions obtained for the three time delay distribution models given in Fig.~\ref{fig:time-delays}, namely $P(t)\propto 1/t$ with $t_{\rm min}=44$ Myr (Model A; blue circles), $P(t)\propto \log(t/t_{\rm min})/t$ with $t_{\rm min}=0.4$ Myr (Model B; orange triangles) and \citet{2018MNRAS.474.2937C} (Model C; green diamonds). The mean delay time is the same in all three models, namely $\langle t_{\rm delay} \rangle=2.37$~Gyr. Observed values are shown by grey crosses. The results for the three models are very similar, except for the lowest metallicity points.}
\label{fig:comp_new}
\end{figure}

\begin{figure}
\centering
\includegraphics[width=0.47\textwidth]{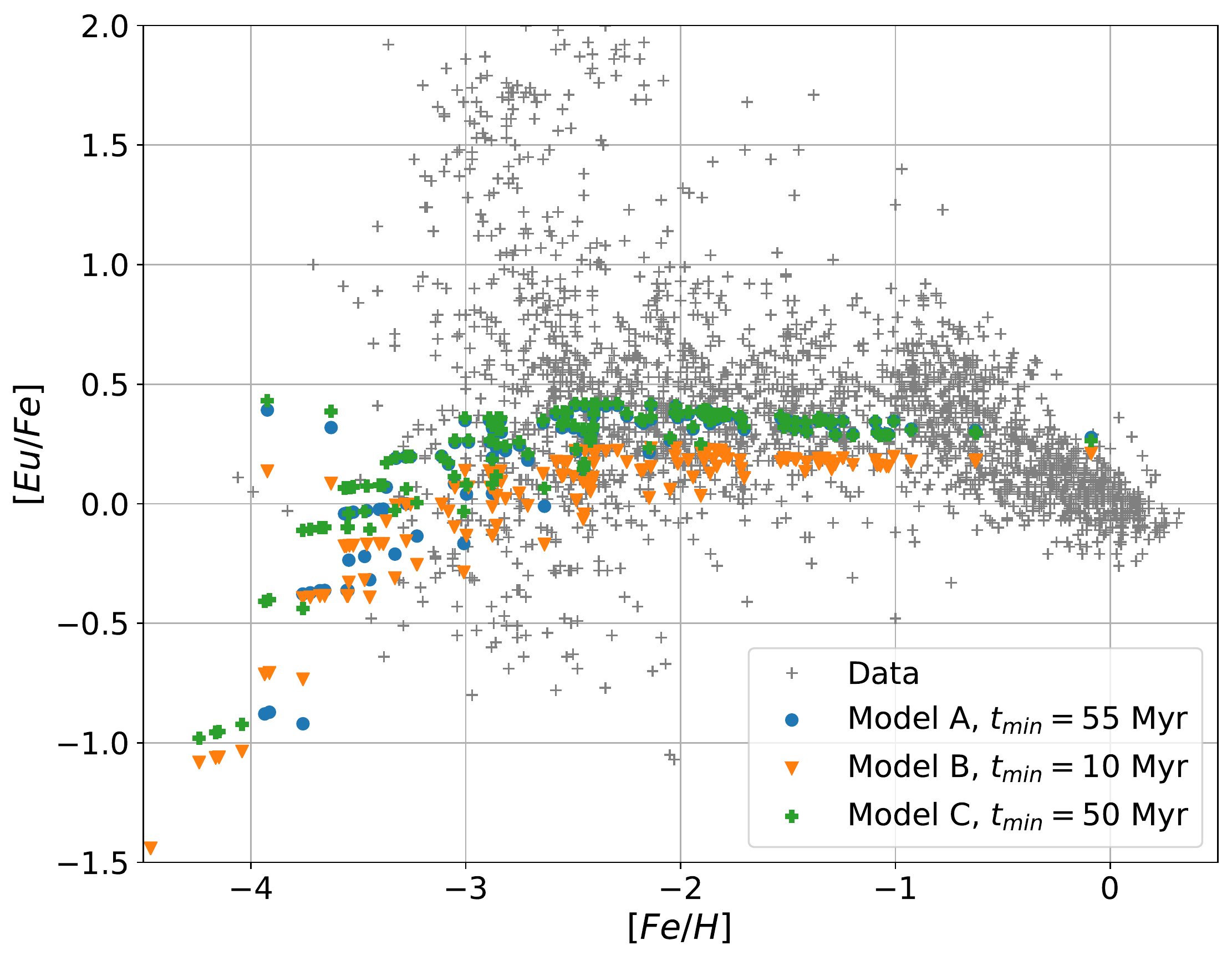}
\caption{Comparison of the three time delay models with $t_{\rm min}$ values chosen to produce the same fraction of rapid mergers $F_{\rm delay}(<100 \rm{Myr})=0.11$: $P(t)\propto 1/t$ with $t_{\rm min}=55$ Myr (Model A; blue circles), $P(t)\propto \log(t/t_{\rm min})/t$ with $t_{\rm min}=10$ Myr (Model B; orange triangles) and \citet{2018MNRAS.474.2937C} (Model C; green diamonds). Observed values are shown by grey crosses. The results for the three models are very similar for the lowest metallicity points, but slightly diverge at the higher metallicity end.}
\label{fig:comp_same_100Myr}
\end{figure}

Finally, in Fig. \ref{fig:comp_new} we compare models A, B and C, keeping $t_{\rm max}=13.8$ Gyr and varying $t_{\rm min}$ so as to obtain the same mean value of $t_{\rm delay}$. It is clear from Fig. \ref{fig:comp_new} that Eu abundance as a function of metallicity depends primarily on this mean value $\langle t_{\rm delay} \rangle$, especially towards higher metallicites. There are however significant differences at the low-metallicity end: Model B appears to be much more efficient in enriching small halos at an early stage. This feature can be traced back to our comparison of the time delay models in Fig.~\ref{fig:time-delays}. As discussed above, for the same mean value $\langle t_{\rm delay} \rangle$, in Model B there are about $27\%$ of mergers with a delay time below $100$ Myr, compared with only $\sim 11\%$ in Model C. These very rapid mergers efficiently enrich the small halos at early times. Indeed, if we choose values of $t_{\rm min}$ that give the same fraction of rapid mergers in all three models, as shown in Fig.~\ref{fig:comp_same_100Myr} (where rapid is defined as a delay below $100$ Myr, {\it i.e.} $F_{\rm delay}(<100 \rm{Myr})=0.11$, see Eq.\ref{eq_frac}), we obtain very similar evolution at low metallicities, as expected. However in this case the models slightly diverge towards higher metallicities, where mergers with long timescales become important.

We also note the evolution of Eu abundance is determined by the product $\alpha_{\rm BNS}m_{\rm Eu}$. We verified that the results remain unchanged if both $\alpha_{\rm BNS}$ and $m_{\rm Eu}$ are varied so as to keep their product constant.

To summarize, we draw two conclusions from Figs. \ref{fig:one_over_t}-\ref{fig:comp_new}. First of all, we observe that our model, which includes BNS mergers as the only sources of r-process elements, reproduces some of the Eu abundances at low metallicities, even though the mean time delays between stellar formation and BNS mergers are of the order of a few Gyrs. This result depends on the BNS merger rate in high-redshift low-mass galaxies, which is at present not constrained observationally (see Fig.~\ref{fig:RNS}). 

Our second major conclusion is that we are not able to reproduce the large observed scatter in the Eu abundance under the assumption of an instantaneous mixing, even by taking into account the formation history of the galaxy from small building blocks. Below we will study this dispersion with an analytic diffusion model.

\section{Europium abundance: turbulent mixing}
\label{sec:inhom}

\begin{figure}
\centering
\includegraphics[width=0.45\textwidth]{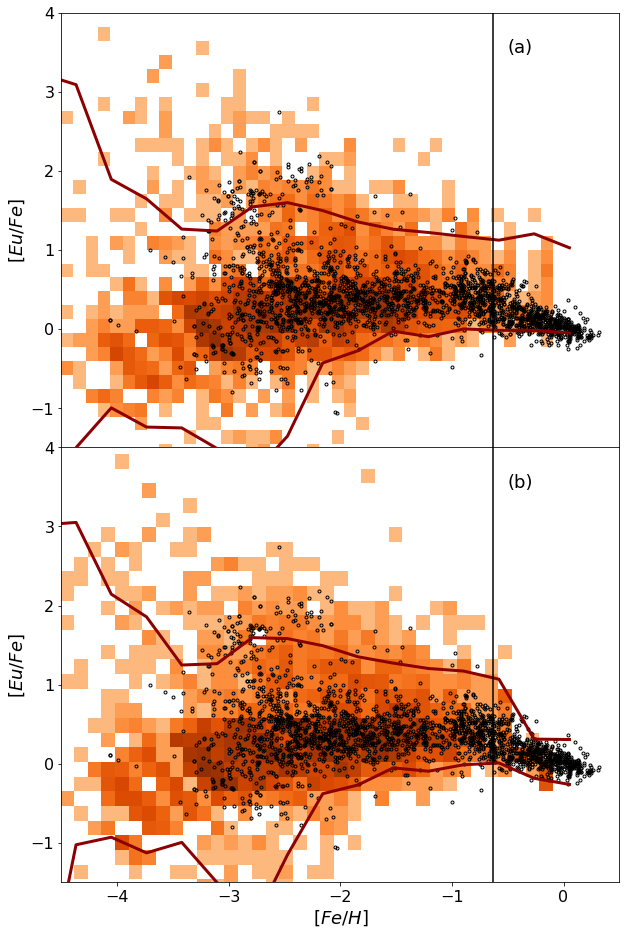}
\caption{Eu abundance in the turbulent mixing model with the $P(t)\propto 1/t$ time delay distribution (Model A) with a minimal time delay of $t_{\rm min}=10$ Myr (see Fig.~\ref{fig:one_over_t}) and diffusion coefficient $D=0.005$~kpc$^2$~Gyr$^{-1}$. The $5\%$ and $95\%$ percentiles are shown by red lines. 
Panel (a): results  with the turbulent mixing scheme as described in \S\ref{sec:inhomogen}.
The model reproduces early enrichment in Eu, but overproduces Eu at high metallicities [Fe/H]$\gtrsim -0.6$. 
Panel (b): same model with
the inclusion of a special treatment of the last branch of the merger tree mimics the evolution of disk stars and results in a better agreement with observations (see text for details). Black vertical lines at [Fe/H]=-0.6 mark the beginning of the last branch. The small differences between the panels at low metallicity are only due to different realizations of the turbulent mixing model.}
\label{fig:mod1-med}
\end{figure}

\begin{figure}
\centering
\includegraphics[width=0.45\textwidth]{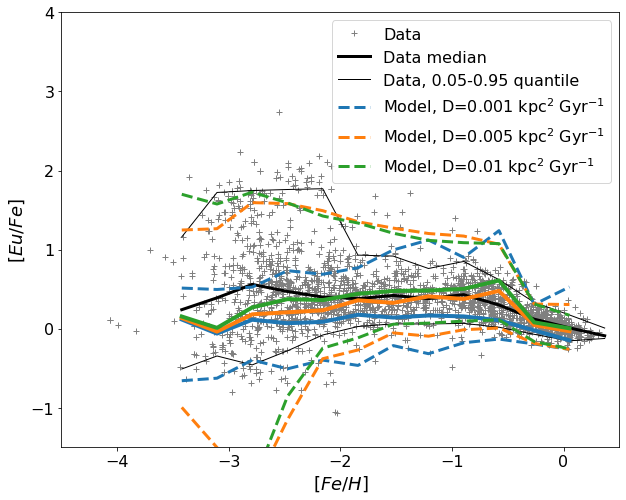}
\caption{Same as Fig.~\ref{fig:mod1-med} with $t_{\rm min}=10$ Myr and varying diffusion coefficients (indicated in the legend). Dotted lines show the $5\%$ and $95\%$ percentiles of each case, while solid lines show the medians. The scatter at low metallicities decreases with $D$ while the median is not affected.}
\label{fig:mod1-varD}
\end{figure}

\begin{figure}
\centering
\includegraphics[width=0.45\textwidth]{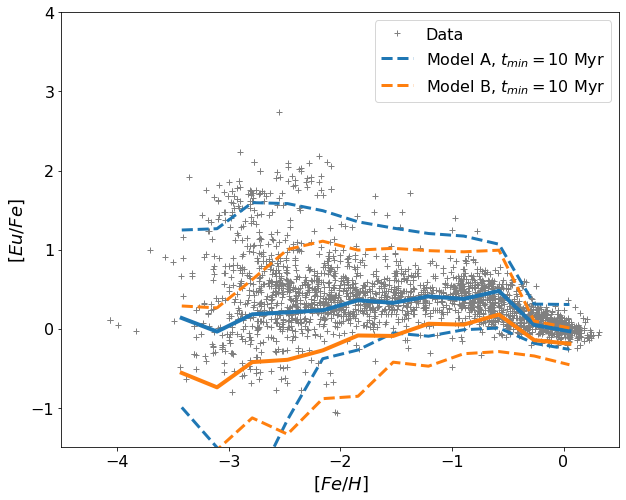}
\caption{Same as Fig.~\ref{fig:mod1-med}, comparing Model A (blue) and Model B (orange) for the delay time distribution. Dashed lines show the $5\%$ and $95\%$ percentiles of each model, while solid lines show the medians. For both models, the diffusion coefficient is $D=0.005$ kpc$^2$~Gyr$^{-1}$ and the minimal delay time is $t_{\rm min}=10$ Myr. Model A is characterized by a mean delay time $\langle t_{\rm delay} \rangle=1.9$~Gyr and Model B by 3.3~Gyr.
}
\label{fig:diff-comp-AB}
\end{figure}

\begin{figure}
\centering
\includegraphics[width=0.45\textwidth]{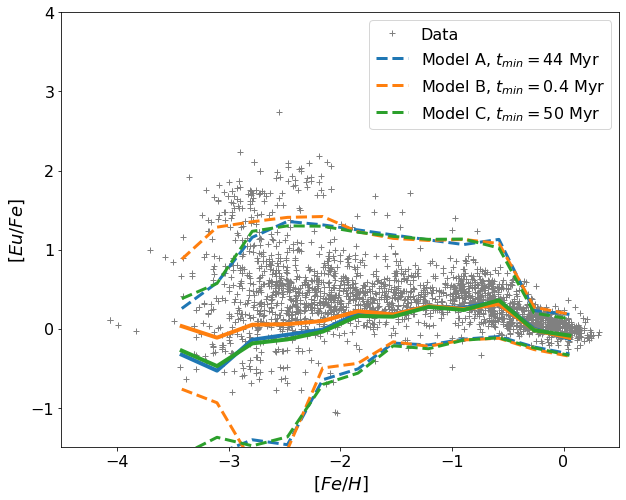}
\caption{Same as Fig. \ref{fig:mod1-med}, comparing predictions based on Model A (blue), B (orange) and C (green) for the time delay distribution, where the respective values for $t_{\rm min}$ are chosen so as to give the same mean delay time $\langle t_{\rm delay} \rangle=2.37$~Gyr (see Fig.~\ref{fig:comp_new}). Dashed lines show the $5\%$ and $95\%$ percentiles of each model. For all models, the diffusion coefficient is $D=0.005$~kpc$^2$~Gyr$^{-1}$.
}
\label{fig:diff-comp-ABC}
\end{figure}

\begin{figure}
\centering
\includegraphics[width=0.45\textwidth]{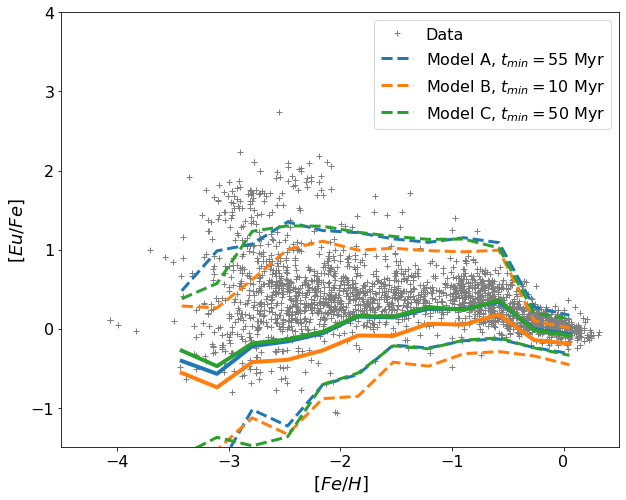}
\caption{Same as Fig. \ref{fig:mod1-med}, comparing predictions based on Model A (blue), B (orange) and C (green) for the time delay distribution, where the respective values for $t_{\rm min}$ are chosen so as to give the same fraction of rapid mergers $F_{\rm delay}(<100 \rm{Myr})=0.11$. Dashed lines show the $5\%$ and $95\%$ percentiles of each model. For all models, the diffusion coefficient is $D=0.005$~kpc$^2$~Gyr$^{-1}$. 
}
\label{fig:diff-comp-ABC-same100Myr}
\end{figure}

In the following we explore the consequences of the turbulent mixing scheme, described in Section \ref{sec:inhomogen}. As discussed above, in this case we obtain multiple values of [Eu/Fe] abundance for each sub-halo, corresponding to the different ``observer'' points (which mimic the stars that form at different locations in the galaxy). For each sub-halo we therefore obtain a distribution of [Eu/Fe] and [Fe/H] values. We bin these values in $20$ metallicity ([Fe/H]) bins of width $0.3$ dex and show the results in 
figures~\ref{fig:mod1-med}--\ref{fig:diff-comp-ABC-same100Myr}.
Note that, in this approach, we have an additional parameter, which is the diffusion coefficient $D$. In the following we explore how this diffusion parameter affects our results.

Since the total number, locations and times of the mergers in each halo are drawn from probability distributions, the evolution of Eu abundance varies for different realizations for a given merger tree. Below we show one realization for each choice of the model parameters, but we checked that these results are representative in that both the median values in each metallicity bin and the percentiles do not change significantly among the different realizations.

As can be seen in Figs.~\ref{fig:one_over_t} and~\ref{fig:mod1-med},
 the median values of our turbulent mixing model do not necessarily coincide with those shown in the previous section within the instantaneous mixing regime. The reason for this is that in the instantaneous mixing case, the average was performed over the whole ISM in the galaxy. In the turbulent mixing scheme, we first calculate the Eu and Fe abundances of our ``observer'' points, and then average over those.

In the upper panel of Fig.~\ref{fig:mod1-med} we show our results for Model A with 
a diffusion coefficient of $D=0.005$~kpc$^2$~Gyr$^{-1}$: the 2D-binned data in the [Eu/Fe]-[Fe/H] plane (orange heatmap) and the $5-95$ percentiles in each bin of [Fe/H] (red lines). We observe that the distribution in our model is in reasonably good agreement with the data up until [Fe/H]$\sim -0.5$. The most interesting feature of this plot are the high [Eu/Fe] values reached even at very low metallicities: these are contributed by observer stars that happen to be born in a recently enriched region. Thus, our semi-analytic model coupled to a turbulent diffusion scheme confirms the conclusions of \citet{2015ApJ...807..115S} obtained with a numerical model of a MW-like galaxy. \citet{2020arXiv200301129B} who introduced this turbulent diffusion scheme, used it in the context of a simplified galaxy model and focused on high metallicities, where they obtained a much higher dispersion than observed. Here we show that this model, when coupled to a detailed galaxy evolution calculation, reproduces the  observed abundances in the low to intermediate metallicity ranges probed by observations. The model predicts extremely Eu-rich (although rare) stars even at [Fe/H]$\sim -4$, however there are very few observations at these metallcities, and in the following we will concentrate on the range [Fe/H]$>-3.5$.

However, the model shown in the upper panel of Fig.~\ref{fig:mod1-med} does not reproduce the decline in [Eu/Fe] observed at high metallicities for stars that reside in the disk. Indeed, there is currently a debate in the literature, since  \citet{2019ApJ...875..106C}  argued that BNS mergers cannot reproduce this observational result and could therefore not be the only source of r-process elements, while \citet{2020arXiv200704442B} recently showed that this problem can be alleviated by including the effect of natal kicks of the NSs. Our model lacks spatial resolution and is therefore not well adapted to explore this population of stars, but we can roughly identify them with the stars born along the last branch of our merger tree i.e. the stars born after the last major merger event, which in our model takes place at the cosmic age of $5.5$ Gyr.
To investigate the impact of the disk geometry, we slightly change our prescriptions after this last major merger event: we
assume that all the stars in the last branch of the merger tree are born in the disk with scale height $h=400$ pc \citep{2016ARA&A..54..529B}. On the other hand, the BNS mergers, even those occurring at later times, can take place anywhere, including the halo, since their progenitor stars formed at much earlier epochs. Therefore, we assume that only BNS that merge in the disk contribute to its chemical enrichment (these BNS could have formed before or after the last major merger). This last assumption limits the number of mergers that contribute to r-process elements in the disk, which leads to the decline in [Eu/Fe] as seen in the lower panel of Fig.~\ref{fig:mod1-med}, in much better agreement with the data at high metallicities. 
Precisely, the transition (last major merger event at $5.5$ Gyr) corresponds to a metallicity of [Fe/H]$=-0.6$ (indicated by the black vertical lines in Fig.~\ref{fig:mod1-med}).
This value slightly depends on our assumptions for the Ia supernova rate, but a lower limit is  [Fe/H]=$-1$, i.e. the contribution of SNII alone to the iron budget at this age.

This geometrical effect is therefore a promising avenue to understand the observed evolution of [Eu/Fe] at high metallicities, although other mechanisms discussed in the literature, such as a metal-dependent efficiency of forming BNS \citep{2019MNRAS.486.2896S}, could lead to a similar effect.
However, we stress that we do not attempt to describe all the observed properties of the disk, such as the differences in the ages and metallicities of MW disk stars  \citep{2017A&A...608L...1H} and their variation as a function of radius \citep{2018ApJ...855..104T}. Clearly, a full spatial model of the different components of the Galaxy is needed in order to describe the chemical evolution of the disk. We also stress that, as can be seen in Fig.~\ref{fig:mod1-med}, the results for metallicities below [Fe/H]$=-0.6$, which are our main subject in this work, are not affected by 
this "disk" prescription, as expected. 
In all the plots below the results are obtained with this new prescription.

Next we explore the effect of the diffusion coefficient~$D$. In Fig.~\ref{fig:mod1-varD}, we show the predictions for our Model~A with $t_{\rm min}=10$ Myr and varying $D$ in the range $0.001-0.01$~kpc$^2$~Gyr$^{-1}$. Both values of $D=0.005$ and $0.01$~kpc$^2$~Gyr$^{-1}$ are in equally good agreement with the data and we conclude that these values are consistent with the observed scatter.

Finally, we study the effect of the BNS merger time delay distribution. In Fig.~\ref{fig:diff-comp-AB} we compare the results for Models A and B, adopting $t_{\rm min}=10$ Myr and $D=0.005$ kpc$^2$~Gyr$^{-1}$ in both cases. We find that the dispersion does not change significantly, strengthening our conclusion that the scatter stems primarily from the diffusion process and therefore essentially depends on the diffusion coefficient $D$. On the other hand, the median value of the relative [Eu/Fe] abundance is lower for Model B, as already discussed above (see Figs.~\ref{fig:one_over_t}-\ref{fig:logt_t}). With a minimal delay time $t_{\rm min}=10$~Myr, Model B is characterized by a mean delay time $\langle t_{\rm delay} \rangle=3.3$~Gyr, significantly larger than the one characterizing Model A of  1.9~Gyr. We conclude that, keeping all the other model parameters fixed, the time delay distribution mainly affects the mean abundances and the threshold metallicity at which the r-enrichment takes place, while the diffusion coefficient $D$ affects the scatter.

In Fig.~\ref{fig:diff-comp-ABC}, we compare Models A, B and C where the respective $t_{\rm min}$ parameter is adjusted so as to give the same value of the mean $\langle t_{\rm delay} \rangle$ (see Fig.~\ref{fig:comp_new}). We observe that there is very little difference between these models except for the lowest metallicity bins, similarly to the instantaneous mixing case shown in Fig.~\ref{fig:comp_new}. Similarly, in Fig.~\ref{fig:diff-comp-ABC-same100Myr} we compare Models A, B and C where the respective $t_{\rm min}$ parameter is adjusted so as to give the same fraction of rapid mergers ($F_{\rm delay}(<100 \rm{Myr})=0.11$; see Fig.~\ref{fig:comp_same_100Myr}). We find that models A, C produce very similar results, but the Eu abundances in Model B are lower because of the larger (on average) merger times. 

As mentioned above, the evolution of Eu abundance in the homogeneous mixing model is determined by the product $\alpha_{\rm BNS}m_{\rm Eu}$. This is also the case for the median value of [Eu/Fe] turbulent mixing model. We find nevertheless that in the lowest metallicity bins there are less Eu-enriched stars if $\alpha_{\rm BNS}$ is lower.

\section{Sensitivity to model parameters}
\label{sec:parstudy}

In this Section we perform a sensitivity study in order to check that our main conclusions are sufficiently robust. We focus on two aspects of the model: dark matter (and stellar) mass and the resolution of the merger tree.

\subsection{Galaxy mass: DM halo and stars}
\label{sec:galmass}

As we discussed above, 
our reference model 
reproduces several important observable features of the Milky Way (\S\ref{sec:evolution}) but 
it has slightly smaller stellar mass and it does not reach super-solar metallicities. One may wonder therefore whether our main results would hold in a model with, for example, a higher stellar mass.

Our fiducial DM merger tree is calculated for a $z=0$ halo of $M_{\rm halo}=10^{12}M_{\odot}$, which corresponds to current estimates \citep{2007MNRAS.379..755S,2016ARA&A..54..529B}. However, there is a significant uncertainty in this value, the different estimates varying by a factor of a few, depending on the technique used. Below we explore two additional models with $M_{\rm halo}=5\cdot 10^{11}M_{\odot}$ and $M_{\rm halo}=2.6\cdot 10^{12}M_{\odot}$, corresponding to the minimal and maximal halo masses from Table 8 of \citet{2016ARA&A..54..529B}. Naturally, the choice of DM halo mass at $z=0$ affects all the other masses in the model, including the mass of stars. Specifically, while our fiducial model reaches a stellar mass of $M_*(z=0)=10^{10}M_{\odot}$, the model with the smallest DM halo reaches $M_*(z=0)=6.3\cdot 10^{9}M_{\odot}$ and the one with the largest DM halo $6.3\cdot 10^{10}M_{\odot}$. Note that the latter case is fully consistent with the stellar mass of the MW. In this latter case we also modified the SFR cutoff mas $M_q$ (see Table~\ref{tab:model_params}) to $M_q=2\cdot 10^{10}M_{\odot}$ to allow for star formation at the later epochs of galaxy evolution.
The parameter $M_q$ used in our SFR law in Eq.~\ref{eq:SFR_law} is a free parameter of the model. In this study we did not attempt to fit a specific star formation history; we therefore leave the determination of this parameter and the resulting uncertainty to future work.

\begin{figure}
\centering
\includegraphics[width=0.45\textwidth]{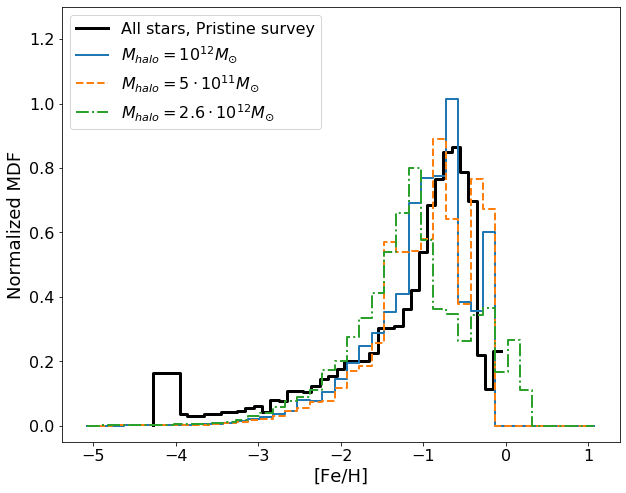}
\caption{MDF with varying DM halo masses at $z=0$ (and different stellar masses, see text). 
Blue: reference model (see Table \ref{tab:model_params}) with $M_{DM}(z=0)=10^{12}M_{\odot},M_*(z=0)=10^{10}M_{\odot}$. 
Orange: $M_{DM}(z=0)=5\cdot 10^{11}M_{\odot},M_*(z=0)=6.3\cdot 10^{9}M_{\odot}$. 
Green: $M_{DM}(z=0)=2.6\cdot 10^{12}M_{\odot},M_*(z=0)=6.3\cdot 10^{10}M_{\odot}$. Data is taken from the Pristine Survey \citep{2020MNRAS.492.4986Y}.
}
\label{fig:comp-SFR-params-MDF}
\end{figure}

\begin{figure}
\centering
\includegraphics[width=0.45\textwidth]{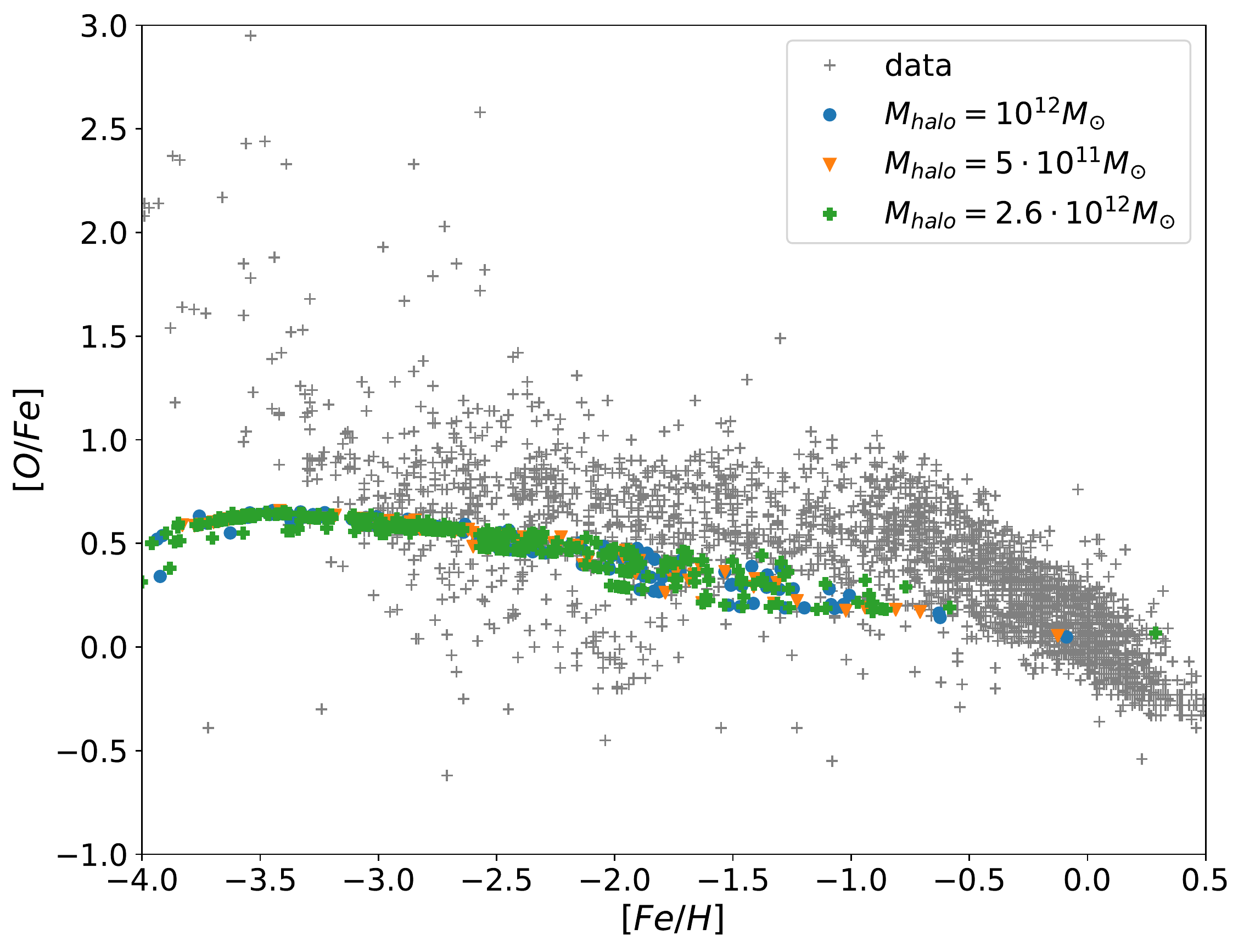}
\caption{Comparison in the [O/Fe]-[Fe/H] plane between models with different DM halos at $z=0$. Blue circles:
 reference model (see Table \ref{tab:model_params}) with $M_{DM}(z=0)=10^{12}M_{\odot},M_*(z=0)=10^{10}M_{\odot}$. 
Orange triangles: 
 $M_{DM}(z=0)=5\cdot 10^{11}M_{\odot},M_*(z=0)=6.3\cdot 10^{9}M_{\odot}$. 
 Green crosses: 
$M_{DM}(z=0)=2.6\cdot 10^{12}M_{\odot},M_*(z=0)=6.3\cdot 10^{10}M_{\odot}$.  
}
\label{fig:comp-SFR-params-O}
\end{figure}

First we explore the metallicity distribution function (MDF) in our three test models. In Fig. \ref{fig:comp-SFR-params-MDF} we compare our results with observational data from the Pristine Survey, that explored the MDF of the MW halo \citep{2020MNRAS.492.4986Y}. Specifically, the data for all stars is taken from their Fig. 1. As can be seen in Fig. \ref{fig:comp-SFR-params-MDF}, all three models are in reasonable agreement with the total observed distribution. Therefore, we conclude that our model is adequate to describe halo stars. However, we caution that none of the models seems to faithfully reproduce the disk stars, as is evident from Fig. \ref{fig:comp-SFR-params-O}, where we plot the oxygen abundance as a function of metallicity: there is no visible change of slope at high metallicities. This difficulty to describe the high-metallicity population is potentially related to the lack of spatial resolution in our model, 
as already mentioned in Section~\ref{sec:inhom}. 

\begin{figure}
\centering
\includegraphics[width=0.45\textwidth]{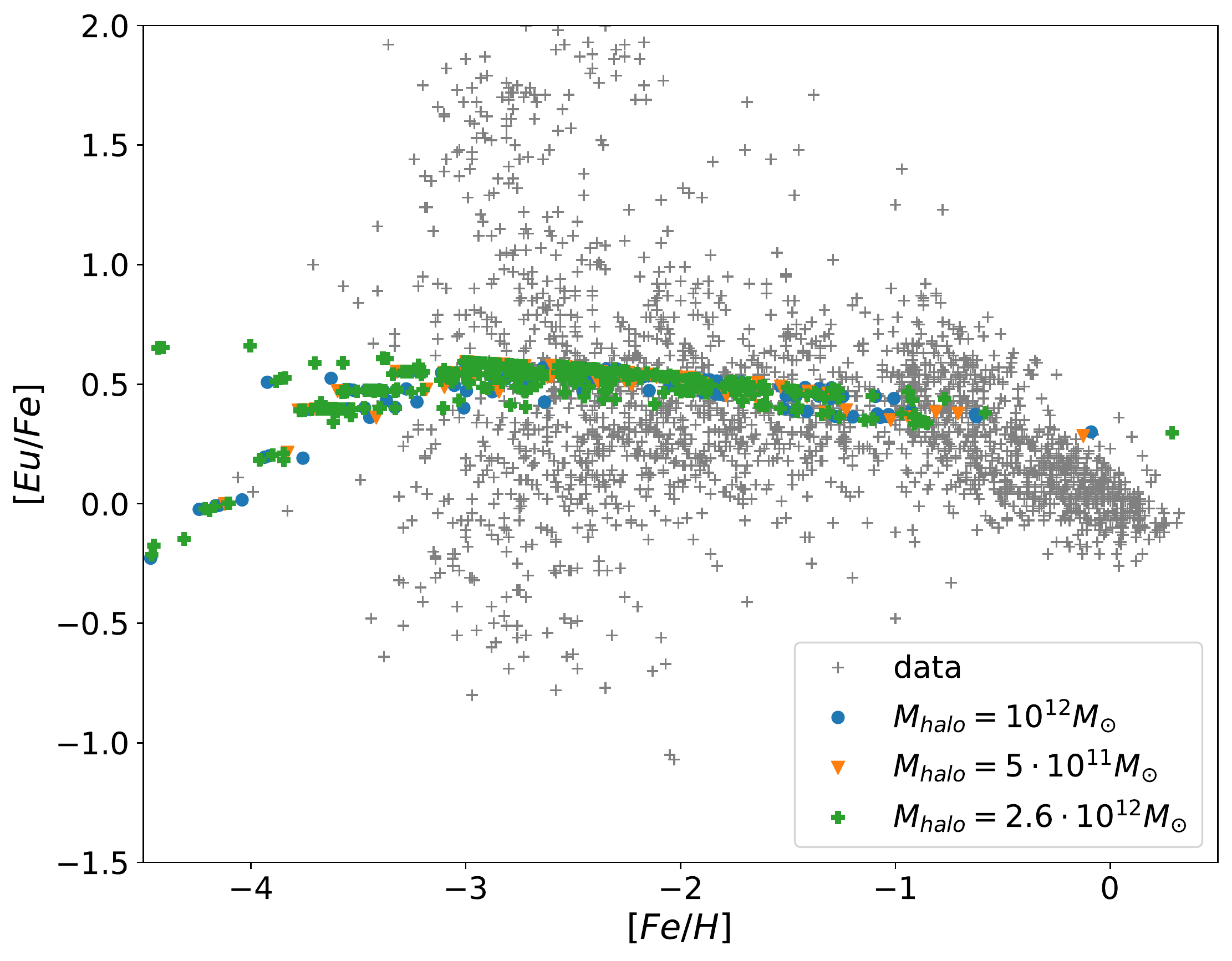}
\caption{Comparison between models with different DM halo mass at $z=0$ (and different stellar masses, see text) for our homogeneous mixing scheme. In all cases we use the time delay distribution $P(t)\propto 1/t$ with $t_{\rm min}=10$ Myr (Model A above).  
}
\label{fig:comp-SFR-params}
\end{figure}

\begin{figure}
\centering
\includegraphics[width=0.45\textwidth]{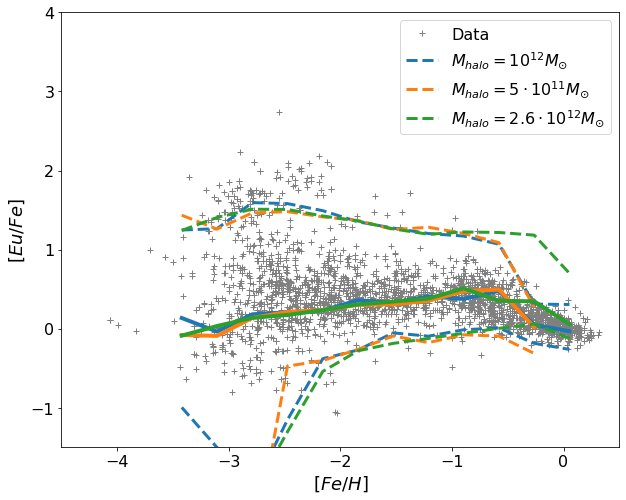}
\caption{Same as Fig. \ref{fig:comp-SFR-params} within the turbulent mixing model.
Solid and dashed lines show the median and $5-95$ percentiles, respectively, for each model.
}
\label{fig:comp-SFR-params-turmixing}
\end{figure}

\begin{figure}
\centering
\includegraphics[width=0.45\textwidth]{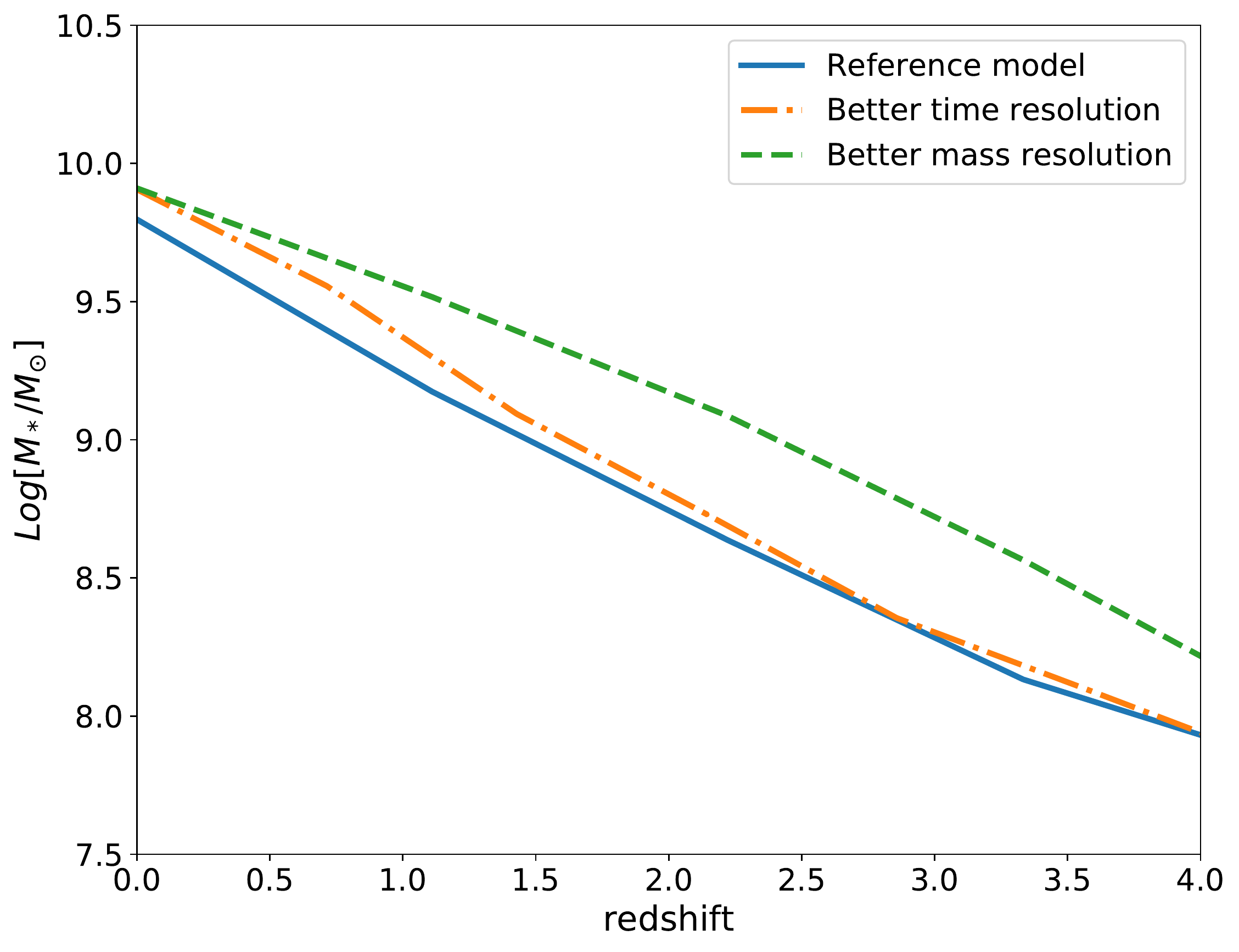}
\caption{Convergence of the stellar mass as a function of the merger tree parameters. 
}
\label{fig:mass-convergence}
\end{figure}

In Figs. \ref{fig:comp-SFR-params} and \ref{fig:comp-SFR-params-turmixing} we compare the abundances in Eu and Fe 
for the three DM halo masses, either assuming our homogeneous or our turbulent mixing scheme.
In both cases the three models produce almost identical results, despite the differences in total stellar masses, as expected for relative abundances. The main difference we observe is in the model with the largest halo mass, which reaches higher metallicities than the other models. From Fig. \ref{fig:comp-SFR-params} we see that the models with halo masses of $5\cdot 10^{11}$, $10^{12}$ and $2.6\cdot 10^{12}M_{\odot}$ reach metallicities of [Fe/H]$=-0.12$, $-0.09$ and $0.29$, respectively.

Based on this analysis we conclude that our main results, in particular the amount of scatter in [Eu/Fe], are not sensitive to the total stellar mass at $z=0$ in our model.

\subsection{Merger tree resolution}
\label{sec:mt_res}

As mentioned above, the fiducial mass resolution for our merger tree was taken to be $M_{\rm res}=10^9M_{\odot}$. While we cannot go to much lower masses for computational reasons, we performed a test where the resolution was reduced to $M_{\rm res}=7\cdot 10^8M_{\odot}$. Another resolution-related parameter is the number of tree levels that are stored for post-processing. As a reminder, we first build the DM merger tree backwards in time with a very small time-step that ensures that we capture all the halos with masses above the resolution mass. However we only store halos at specific output times, specifically at $N_{\rm level}=19$ for our fiducial model. We then evolve the model forward in time, adding gas to the halos and solving the equations for gas inflow/outflow, SFR, etc.,
as described in Section \ref{sec:evolution}.  These evolution equations are solved ``along the branch'' between the times of saved output. When we reach the time of saved output, that means that a merger occurred, the smaller halos fall into the bigger one and we assume the ISM is immediately mixed. Therefore, increasing the number of saved levels $N_{\rm level}$ would, on the one hand, allow a more detailed description of all the subhaloes, but on the other hand introduce a more frequent mixing on the scale of each galaxy. To test the influence of our assumptions we ran a model with $N_{\rm level}=29$ levels.

\begin{figure}
\centering
\includegraphics[width=0.45\textwidth]{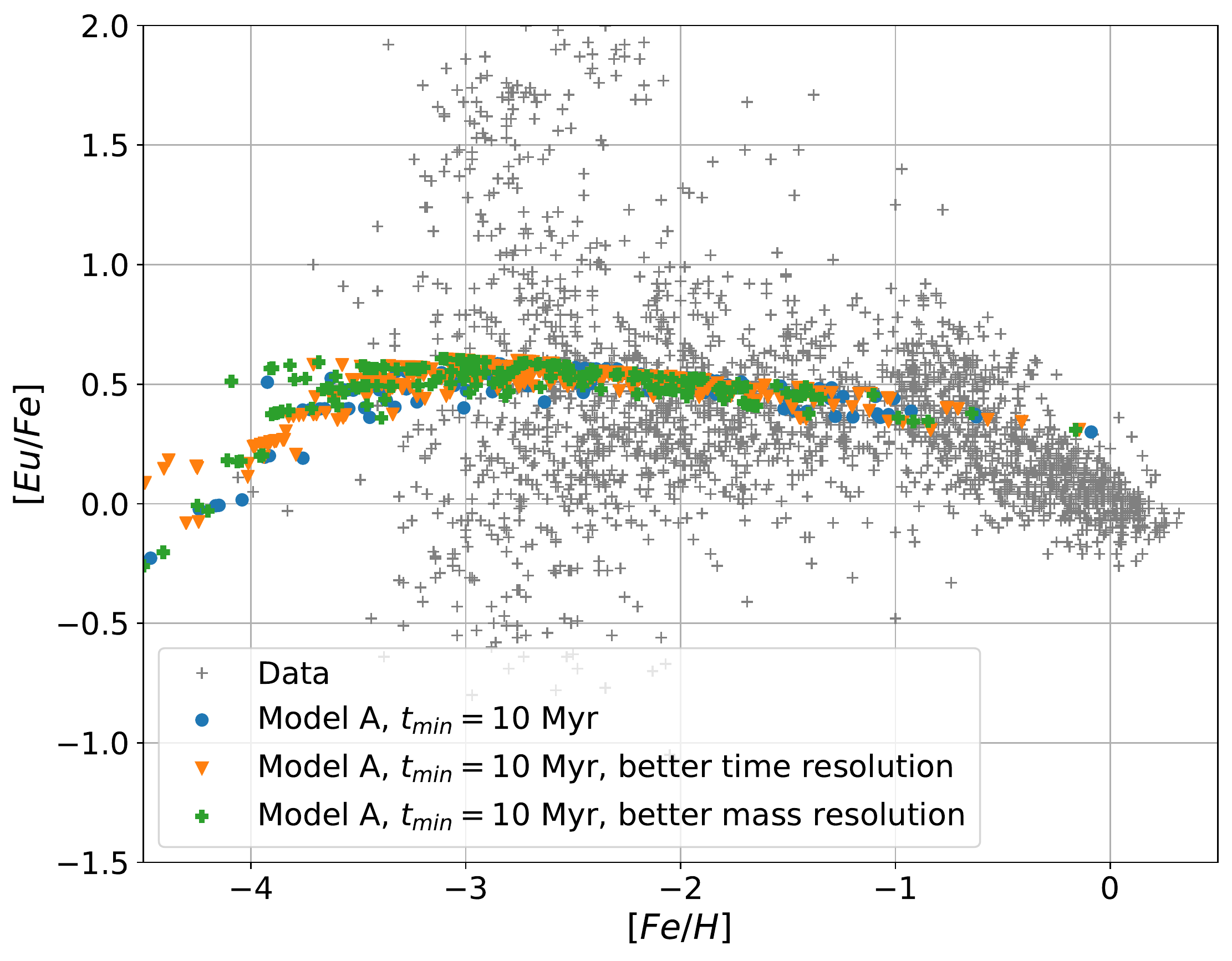}
\caption{Convergence in the [Eu/Fe]-[Fe/H] plane as a function of the merger tree parameters: homogeneous mixing. In all cases we use the time delay distribution $P(t)\propto 1/t$ with $t_{\rm min}=10$ Myr (Model A above). Blue circles: reference model with $19$ tree levels, $M_{\rm res}=10^9 M_{\odot}$. Orange triangles: $29$ tree levels. Green crosses: $M_{\rm res}=7\cdot 10^8 M_{\odot}$.   
}
\label{fig:comp-resolution}
\end{figure}

\begin{figure}
\centering
\includegraphics[width=0.45\textwidth]{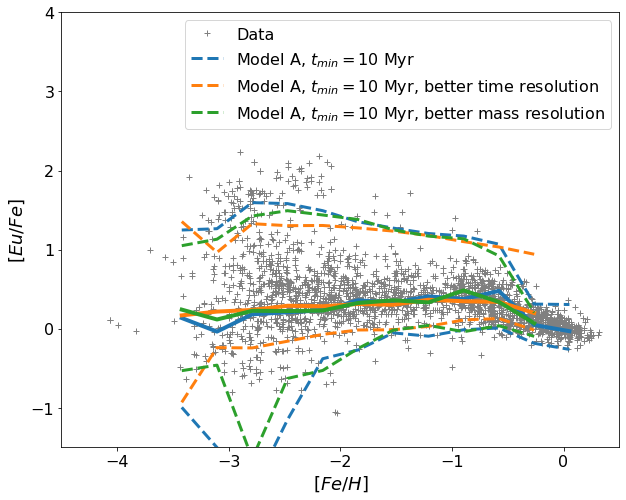}
\caption{Same as Fig.~\ref{fig:comp-resolution} within the turbulent mixing model. 
Solid and dashed lines show the median and $5-95$ percentiles, respectively, for each model. 
}
\label{fig:comp-resolution-turmixing}
\end{figure}

The results for these resolution tests are shown in Figs. \ref{fig:mass-convergence}, \ref{fig:comp-resolution} and \ref{fig:comp-resolution-turmixing}. While the stellar mass differs slightly among these models (within a factor $\sim 3$), it converges to the same value by $z=0$, and the abundances of Eu and Fe are not at all sensitive to these changes. In particular, there is no difference between the models with $N_{\rm level}=19$ vs. $N_{\rm level}=29$ merger tree levels.

\section{Discussion}
\label{sec:discussion}

The results presented above suggest that BNS mergers can be the dominant r-process producing site, and in particular our model can reproduce both the mean value and the scatter in [Eu/Fe] observed in low to intermediate metallicity stars in the MW, while a very crude ``disk'' prescription
allows also to reproduce the high-metallicity part.
We also showed that the two key model parameters that govern this evolution are the time delay distribution of BNS mergers and the diffusion coefficient.
These conclusions are in contrast with some of the previous studies, according to which BNS mergers are not able to produce r-process elements in sufficient amounts at early times \citep{2004A&A...416..997A,2015MNRAS.447..140V, 2018MNRAS.477.1206N, 2018ApJ...855...99C,2020MNRAS.494.4867V}, but in qualitative agreement with \citet{2015ApJ...807..115S,2016ApJ...830...76K,2017MNRAS.466.2474H}.

What could be the source of this discrepancy between different models? As a representative case, we compare our model with  \citet{2020MNRAS.494.4867V}. In this study, the authors use an hydrodynamical model of a MW-like galaxy and a phenomenological model for BNS mergers. To calculate the BNS merger rate they define a constant BNS merger efficiency per unit of stellar mass formed of $3\times 10^{-6}M_{\odot}^{-1}$. In our model the equivalent quantity $R_\mathrm{merger}/\Psi$ is not constant (see Eqs.~\ref{eq:SFR_law} and~\ref{eq:Rmerger}), but varies with redshift and among sub-halos, ranging from $10^{-7}M_{\odot}^{-1}$ to $10^{-4}M_{\odot}^{-1}$. We also observe that in our model this quantity increases with metallicity, but at a given metallicity the scatter reaches about two orders of magnitudes. \citet{2020MNRAS.494.4867V} define a time delay distribution for the BNS mergers with $P(t_{\rm delay})\propto t_{\rm delay}^{-1}$, as in our Model A, with $t_{\rm min}=30$~Myr. The merger rate at $z=0$ for their fiducial model is $1.7~10^2$~Myr$^{-1}$ at the high end of the value inferred from Galactic double NSs, whereas we obtain $10$ Myr$^{-1}$, slightly lower than the low end observational value.  We also note that we use $m_{\rm Eu}=10^{-4} M_{\odot}$ for the mass of Eu produced in each BNS merger, while \citet{2020MNRAS.494.4867V} use $m_{\rm Eu}=4.7~10^{-5} M_{\odot}$ 
(but note that the mean evolution in our model is only determined by the product $\alpha_\mathrm{BNS}\, m_\mathrm{Eu}$).
Another key difference between our models is the SFR, which in our model is always slightly below the one in the MW, while \citet{2020MNRAS.494.4867V} use values a factor of a few above the measured values in the MW. 

To conclude, we listed above the main potential sources of differences, but a comprehensive comparison between the different models present in the literature is beyond the scope of the present study. We suggest that the differences are manifested in the merger rate of BNS in small, low-metallicity galaxies, and that r-process abundances can indeed be used as an indirect probe of these environments, but we leave a more complete analysis to future work.

An additional observational feature that could help distinguish between the different scenarios is the decrease in the scatter in [Eu/Fe] towards intermediate metallicities \citep{2015A&A...577A.139C}. This effect could be caused by the fact that sources with different lifetimes (fast/slow BNS mergers and rare SNe) contribute to the r-process enrichment.

\section{Conclusions}
\label{sec:concl}

The question of the origin of r-process elements, while witnessing major advances in the past few years, still remains open. On the one hand, the spectacular multi-messenger observations of the coalescing BNS 170817 \citep{2017PhRvL.119p1101A,2017ApJ...848L..12A,2017Sci...358.1556C,2017ApJ...848L..12A,2017ApJ...848L..19C} has confirmed the existence of kilonovae and the associated production of r-process elements in such systems. On the other hand, chemical evolution models struggle to reproduce the r-process abundance observed in MW stars, both ultra-metal-poor stars as well as young disk stars.

In this paper, we studied the abundance of r-process elements with a semi-analytic galactic chemical evolution model, focusing on BNS mergers. We explored several assumptions regarding the time delay distribution of BNS mergers, and implemented a turbulent mixing scheme to study the dispersion of r-process elements in the ISM. Our main conclusions can be summarized as follows:

\begin{itemize}
    \item BNS mergers can be a dominant r-process producing site, and in particular both the merger rates and time delays predicted by population synthesis models are compatible with the Eu abundances observed in MW stars.
    \item The assembly history of our model galaxy from small building blocks has only a weak impact on the scatter in the Eu abundances.
    \item Turbulent mixing of the freshly synthesized elements in the ISM leads to a significant dispersion in the [Eu/Fe]-[Fe/H] plane at low metallicities, in good agreement with the observed scatter for a moderate diffusion coefficient.
\end{itemize}

We investigated several functional forms of the BNS merger time delay distribution, and found that the shape of the distribution at the lower end plays a critical role in the resulting r-process abundances in low-mass low-metallicity halos. This result underlines the importance of studying fast-merging BNS scenarios with population synthesis models. More multi-messenger BNS detections may provide new constraints in the future on such a fast-merging population as the afterglow should be easier to detect with the denser close environment expected in this case \citep{2020A&A...639A..15D}.

We also wish to stress several caveats to our model, that may have an important effect on the final results. First of all, we do not take into account the birth kicks of the neutron stars. Indeed, newly-born neutron stars can acquire velocities of up to hundreds of km/sec. As a result, not only the binary may be potentially disrupted, which is already taken into account in the low value of $\alpha_\mathrm{BNS}$, but even if it survives it may find itself far away from its birth site. The r-process elements produced during the merger will therefore be released in a metallicity environment very different from the one the binary itself was born in \citep{2020arXiv200704442B}. However, we expect that this effect will not have a strong influence on the fast-merging BNS that enrich the low-metallicity environments \citep{2019MNRAS.486.3213A}.

The second important caveat concerns the mass released in r-process elements in each BNS merger event. In this work we assumed that the NS masses were constant, and each BNS merger produced the same mass in Eu.
By comparing the kilonova associated to GW170817 and a sample of short GRB afterglow lightcurves,  \citet{2018ApJ...860...62G} find some evidence for diversity in the emission properties of kilonovae that could not be explained by the viewing angle alone. This could point  to differences in the ejected mass, which depends on the masses of the binary components \citep{2019ARNPS..69...41S}. Diversity is especially expected in the case of BH-NS mergers, the contribution of which is neglected in the present study\citep[see the discussion in ][]{2016MNRAS.455...17V}, as their rates are expected to be    significantly smaller than those of BNS \citep{2019PhRvX...9c1040A}.

We do not include the
interactions of a central massive black hole with the surrounding gas 
which
can cause galactic outflows and affect the abundances in the ISM. This effect can be important, especially in view of the growing evidence that feedback from central black holes can play a role in dwarf galaxies \citep{2020arXiv200710342K}.

Finally, we emphasize that we did not attempt to produce an accurate model of the MW, and indeed our model galaxy has lower stellar mass and fewer high-metallicity stars than the MW. 
However, as discussed in Section~\ref{sec:parstudy}, our main results, and in particular the amount of scatter in [Eu/Fe], are not sensitive to the stellar mass at $z=0$. 
Better understanding the precise MDF of the Milky Way and specifically the enrichment of disk stars would require a more detailed treatment of the various Galactic components where our model lacks spatial resolution. This is also the case for understanding the observed decreasing trend of [Eu/Fe] with [Fe/H] at high metallicity, which are mainly located in the disk. To illustrate this aspect, we included in our model a simple prescription to restrict the turbulent mixing in the disk at late epochs and managed to reproduce the observed evolution of [Eu/Fe] at [Fe/H]$>-0.6$.

In view of these caveats we conclude that the role of BNS mergers as sources of r-process elements may be more prominent than according to other works, but is still not entirely clear. In future work we plan to include other possible sources
associated to rare types of core-collapses.
Moreover, further progress in the modeling of fast-merging BNS is needed in order to fully understand r-process abundances in low-metallicity environments.

\section*{Acknowledgments}
We thank the referee Gabriele Cescutti for constructive comments that helped to improve the manuscript. The authors acknowledge support from the CNRS-IRP project "Origin of heavy elements in the Universe: compact stars and nucleosynthesis". I.D. acknowledges support from the "Tremplins nouveaux entrants" program of Sorbonne Universit\'{e}. F.D., I.D. and E.V. acknowledge financial support from the Centre National d'\'Etudes Spatiales (CNES).
S.G. acknowledges financial support from FNRS (Belgium). This work was partially supported by the Fonds de la Recherche Scientifique - FNRS and the Fonds Wetenschappelijk Onderzoek - Vlaanderen (FWO) under the EOS Project No O022818F.

\section*{Data Availability Statements}
The data underlying this article will be shared on reasonable request to the corresponding author.

\bibliographystyle{mn2e}
\bibliography{bns}
\label{lastpage}

\end{document}